\documentclass[aps,pre,preprint,groupedaddress]{revtex4-1}

\usepackage{graphics}
\usepackage{graphicx}
\usepackage{amsfonts}
\usepackage{textcomp}
\usepackage{amssymb}
\usepackage{amsmath}
\usepackage{color}
\usepackage{ulem}
\usepackage{multirow}
\usepackage{units}

\usepackage{soul}

\begin{document}

\title{Surface correlation functions of dead-leaves models}

\author{Cedric J. Gommes$^{1}$} \email{cedric.gommes@uliege.be}

\affiliation{$^1$Department of Chemical Engineering, University of Li\`ege, Li\`ege 4000, Belgium}

\date{\today}

\begin{abstract}
The pore-surface and surface-surface correlation functions are structural characteristics that play an important role in theoretical materials science and in small-angle scattering theory. Exact analytical expressions for the surface correlation functions are available only for very few models, and we here derive such expressions for the general class of dead-leaves models. Within these models, a two-phase pore/solid structure is created by sequentially and randomly filling space with pore-like or solid-like grains that overlap any pre-existing structure, in the same way as dead leaves fall on the ground. The obtained mathematical expressions are valid for any grain shape, in arbitrary dimension. The results are illustrated with monodispersed spherical grains, as well as with a dead-leaves realization of a Debye random medium. In the latter case, the size distribution of the grains is designed to produce a structure having exponential two-point correlation function. Compared to Debye random media obtained by numerical reconstruction, the dead-leaves structure has almost identical surface-surface correlation function, but distinctly different pore-surface correlation function. As a byproduct of our analysis, we also submit a general expression for the pore-surface and surface-surface correlation functions of the Boolean model, valid for arbitrary grains.
\end{abstract}

%\pacs{05.20.-y, 61.43.-j}

\maketitle

\section{Introduction}

Many materials, natural or synthetic, have disordered microstructures. Porous materials, composites, micro-phase separated polymers, alloys, microemulsions, are a few examples among many others. In all these cases, characterizing the microstructure and understanding how it determines the materials' macroscopic properties, is conveniently done using probabilistic concepts \cite{Matheron:1967,Serra:1982,Ohser:2000,Torquato:2002,Lantuejoul:2002}. 

A classical example is the two-point correlation function, also referred to as the covariance, which is defined as the probability for two randomly-chosen points to belong to specific phases of the structure. Two-point correlation functions are central to the elastic and inelastic small-angle scattering properties of nanostructures \cite{Debye:1957,Lindner:2024} as well as to understanding effective elastic and transport properties of composites and porous materials \cite{Torquato:2002}. A two-point correlation function, however, is generally an incomplete characterization of a structure \cite{Gommes:2012}. A more thorough description of a disordered structure therefore requires additional structural descriptors, such as higher-order correlation functions \cite{Matheron:1967,Torquato:2002}, opening granulometries \cite{Serra:1982}, two-point cluster functions \cite{Jiao:2009}, chord-length distributions \cite{Mering:1968,Lu:1993,Roberts:1999,Gommes:2020}, pore-size functions \cite{Torquato:2002,Skolnick:2021}, etc. Each of these functions capture specific characteristics of the structure that other descriptors are blind to. 

The present contribution is concerned with the pore-surface and surface-surface correlation functions, the definitions of which involve the probabilities for random points to belong not just to specific phases, but also to interfaces \cite{Doi:1976,Teubner:1990,Torquato:2002}. These functions play an important role in the calculation of rigorous bounds on the permeability of porous materials \cite{Doi:1976}, as well on survival times in diffusive processes \cite{Torquato:2002}. They are also important in the context of small-angle scattering, notably for micro-emulsion in film-contrast experiments \cite{Teubner:1990,Gommes:2021}. Surface correlation functions also play a role in the context of neutron scattering, when the conditions for the Born approximation of scattering theory are not met \cite{Frielinghaus:2025}. 

Many works have been published about numerical methods for measuring surface correlation functions on discretized structures obtained by simulation \cite{Seaton:1986,Ma:2018}, reconstruction \cite{Jiao:2009,Ma:2020}, or imaging \cite{Samarin:2023,Postnicov:2024}. By comparison, analytical expressions concerning exact values for specific structural models appear to be more limited. The pore-surface and surface-surface functions are well-known for the Boolean model of interpenetrating spheres \cite{Doi:1976}. They have been generalized to arbitrary dimension \cite{Torquato:2002}, and more compact expressions have been proposed for interpenetrating disks \cite{Ma:2020,Samarin:2023}. The analytical results available for other models are sparser.  For Gaussian excursion sets, an exact analytical expression for the pore-surface correlation function is available only in the one-dimensional case, which was obtained as an intermediate step in the calculation of chord length distributions \cite{Roberts:1999}. In the three-dimensional case, only approximations are available which have been proposed for small-angle scattering data analysis \cite{Gommes:2019}, and later generalized to higher-order surface correlations \cite{Cherkasov:2026}. Semi-empirical expressions have also been proposed for numerically-reconstructed Debye random media, {\it i.e.} for structures having exponential two-point correlation function \cite{Ma:2020}. Here, we derive analytical expressions for the pore-surface and surface-surface correlation functions of the dead-leaves model. The results are valid for any grains, in arbitrary dimension. 

The first section of the paper is devoted to a few definitions and some background results. This is followed by the discussion of the dead-leaves model. The formalism used for calculating the surface correlation functions is recursive, and differs from the usual discussion of the dead-leaves model. Some classical results are therefore presented first, including the two-point correlation function and surface area of the dead-leaves models. The recursive formalism is introduced afterwards; we show how the classical results are obtained in that different formalism, and we derive the expressions of the surface correlation functions. The results are afterwards discussed, based also on simulations of three-dimensional dead-leaves models with different size distributions. The paper has two appendices.

\section{Two-point and surface correlation functions}

We refer hereafter to the two phases of the material as the pores and the solid, which we label $0$ and $1$, but the discussion holds for any two-phase structure. The structure is comprehensively characterized by the indicator function of one of the phases, say \cite{Matheron:1967,Torquato:2002}
\begin{eqnarray}
\mathcal{I}_{0}(\mathbf{x}) = \left\{
\begin{array}{ll}
1 & \textrm{if } \mathbf{x} \in \textrm{phase } 0 \cr
0 & \textrm{otherwise} 
\end{array}
\right.
\end{eqnarray}
The indicator function of the other phase is $\mathcal{I}_{1}(\mathbf{x}) = 1 -\mathcal{I}_{0}(\mathbf{x})$, because they are complementary to each other.  In case of stochastic models, the values of the indicator functions are random variables, and we refer to average values through brackets $\langle \rangle$. These can ensemble averages calculated over a large number of independent realizations, or spatial averages calculated over all possible positions in a given realization. The two types of averages are mathematically equivalent for the scope of the present paper.

The porosity is defined as the average of the pore indicator function
\begin{equation} \label{eq:phi_def}
\phi_0 = \langle \mathcal{I}_{0}(\mathbf{x}) \rangle
\end{equation}
and the solid fraction is $\phi_1=1-\phi_0$. The pore-pore two-point correlation function is defined as
\begin{equation} \label{eq:C00_def}
C_{00}(r) = \langle \mathcal{I}_{0}(\mathbf{x}) \mathcal{I}_{0}(\mathbf{x} + \mathbf{r})  \rangle
\end{equation}
{\it i.e.} as the probability for two points at distance $r$ from one another to jointly belong to the pores. Here we consider statistically isotropic structures, for which the dependence is only through the modulus $r = |\mathbf{r}|$. For any statistically isotropic structure, the small-$r$ behavior of the correlation function satisfies
\begin{equation} \label{eq:C00_small_r}
C_{00}(r) = \phi_0 - \frac{s}{4} r + \ldots
\end{equation} 
where $s$ is the specific surface area \cite{Debye:1957}. And for large distances $r$, in absence of long-range order, the value of $C_{00}(r)$ converges to $\phi_0^2$ because two distant points belong to the pores independently of one another. 

The pore-surface correlation function is concerned with the probability for two points at distance $r$ from another to belong to the pore and to the surface, respectively \cite{Teubner:1990,Torquato:2002,Ma:2018}. Because the probability for a point to be on a surface is strictly zero, this interpretation of has to be understood as a limit. For that purpose, we define $\mathcal{S}_\epsilon(\mathbf{x})$ to be the indicator function of a layer with uniform thickness $\epsilon$ containing the surface. The two-point cross-correlation between the pore and the surface layer is defined as
\begin{equation} \label{eq:C0S_def}
C_{0S_\epsilon}(r) = \langle \mathcal{I}_0(\mathbf{x}) \mathcal{S}_\epsilon (\mathbf{x} + \mathbf{r}) \rangle
\end{equation}
The pore-surface correlation is then defined as the limit \cite{Teubner:1990}
\begin{equation}  \label{eq:F0s_limit}
F_{0s}(r)= \lim_{\epsilon \to 0} \frac{C_{0S_\epsilon}(r)}{\epsilon}
\end{equation}
As we discuss in more detail later, the exact position of the pore-solid interface inside the surface layer $S_\epsilon$ is inconsequential in the limit of vanishingly small thickness $\epsilon$, for any finite $r$. 

In the case of structures with no long-range order, the pore-surface correlation function satisfies the following limit \cite{Torquato:2002}
\begin{equation} \label{eq:F0s_large_r}
\lim_{r \to \infty }F_{0s}(r) = \phi_0 s
\end{equation}
and for three-dimensional structures the short-distance behavior satisfies \cite{Teubner:1990}
\begin{equation} \label{eq:F0s_small_r}
F_{0s}(r) \simeq \frac{s}{2} \left[ 1 - \frac{\bar H}{2} r + \ldots \right]
\end{equation}
where $\bar H$ is the surface-averaged mean curvature, defined as $(1/R_1+1/R_2)/2$ where $R_1$ and $R_2$ are the principal radii of curvature at the considered point. The sign of $\bar H$ in Eq. (\ref{eq:F0s_small_r}) is a matter of convention. Here we count the mean curvature positively for a pore/solid interface that is convex towards the solid. In the case of surfaces containing singularities ({\it e.g.} sharp edges) the function $F_{0s}(r)$ still depends linearly on $r$ for small distances but the slope is not directly related to the mean curvature \cite{Ma:2018}. To avoid any ambiguity, we refer to $\bar H$ as the {\it apparent} curvature whenever the distinction is necessary.

In the case of structures that have phase-inversion symmetry, the cross-correlation between the pore and the surface layer $C_{0S_\epsilon}(r)$ defined in Eq. (\ref{eq:C0S_def}) is identical to the cross-correlation between the solid and the surface layer $C_{1S_\epsilon}(r)$. Because the two functions add up to the volume fraction of the surface layer $\phi_{S_\epsilon} = s \epsilon$, one immediately concludes that 
\begin{equation}
F_{0s}(r) = F_{1s}(r) = s/2
\end{equation}
for all $r$, in case of a phase-symmetric structure \cite{Teubner:1990}.

The surface-surface correlation function is concerned with probabilities for two points at distance $r$ from one another to both lie on the surface \cite{Teubner:1990,Torquato:2002,Ma:2018}. Similarly to the pore-surface correlation function, however, $F_{ss}(r)$ is strictly defined as the following limit \cite{Teubner:1990}
\begin{equation} \label{eq:Fss_limit}
F_{ss}(r) = \lim_{\epsilon \to 0} \frac{C_{S_\epsilon S_\epsilon}(r)}{\epsilon^2}
\end{equation}
where
\begin{equation} \label{eq:CSS_def}
C_{S_\epsilon S_\epsilon}(r) = \langle \mathcal{S}_{\epsilon}(\mathbf{x}) \mathcal{S}_{\epsilon}(\mathbf{x} + \mathbf{r} ) \rangle 
\end{equation}
is the two-point correlation function of the same $\epsilon$-thin surface layer as used in Eq. (\ref{eq:C0S_def}). In the limit of small distances $r$, the surface correlation function of a three-dimensional structure has the following general asymptotic dependence \cite{Teubner:1990,Ma:2018,Ma:2020}
\begin{equation}  \label{eq:Fss_small_r}
F_{ss}(r) \simeq \frac{s}{2 r}
\end{equation}
and for large $r$, the asymptotic value is
\begin{equation} \label{eq:Fss_large_r}
\lim_{r \to \infty }F_{ss}(r) = s^2 
\end{equation}
in absence of long-range order. 

\section{The dead-leaves model}

\subsection{Classical definition and results}

\begin{figure}
\includegraphics[width=8cm]{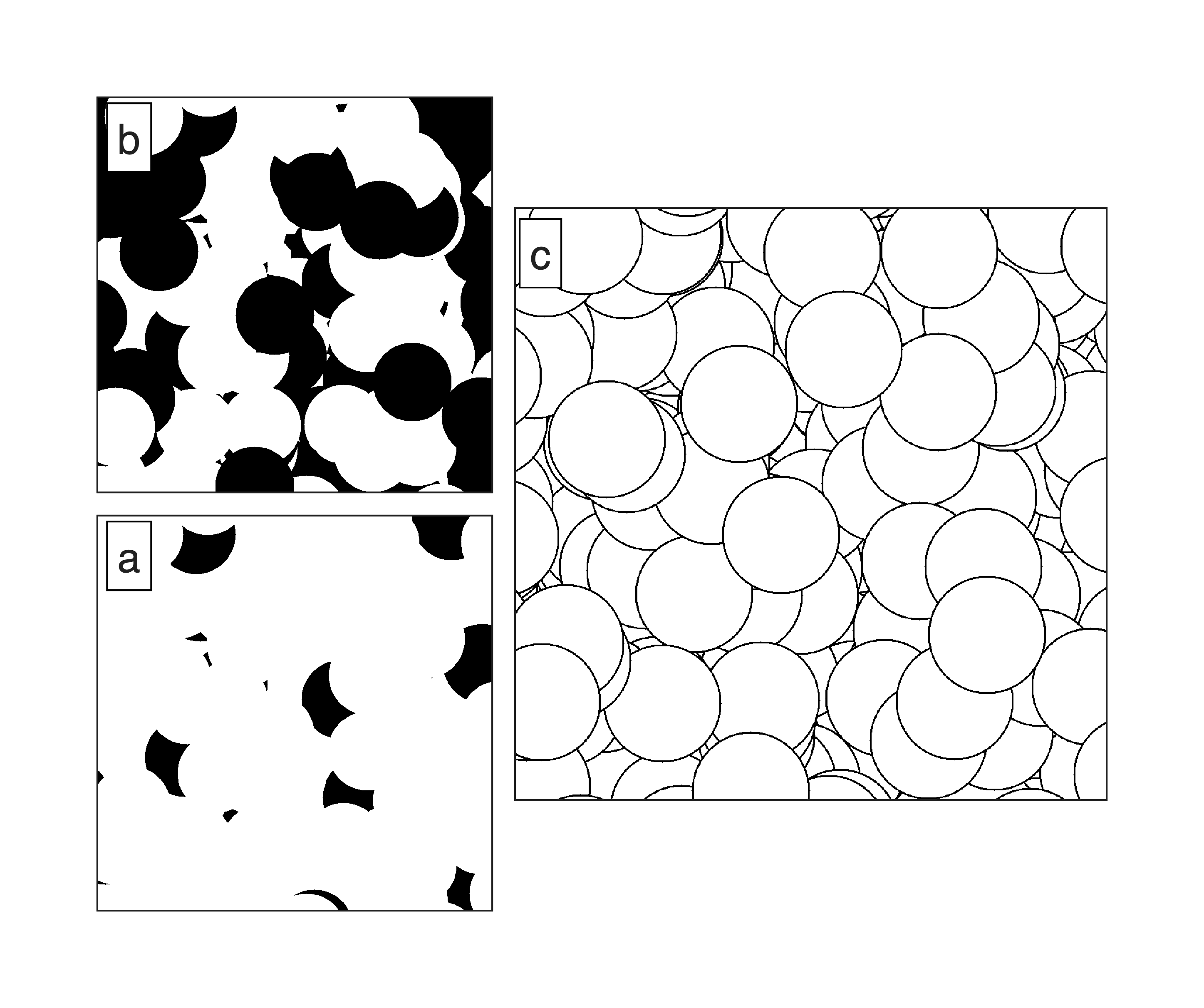}
\caption{Two-dimensional dead-leaves model, whereby pore-like (black) or solid-like (white) disks are sequentially and randomly added to the plane, thereby covering any preexisting structure. Two fractions of pore-like disks are considered $\phi_0= 0.1$ (a) and $\phi_0=0.5$ (b). The construction is equivalent to using random disks to construct a tessellation (c), each cell of which is afterwards randomly assigned to the pores or solid.}
\label{fig:deadleaves}
\end{figure}

The dead-leaves model is equivalent to a time-dependent Boolean model of overlapping grains \cite{Matheron:1968B}. Randomly positioned grains $G$ centered on a Poisson point process are deposited one at a time, and each new grain covers any pre-existing structure. In the limit of asymptotically large deposition times, the statistical properties of the structure converge to well-defined values, which only depend on the grain characteristics. The model is illustrated in Fig. \ref{fig:deadleaves}a, for the two-dimensional case of circular grains that are solid-like or pore-like with identical probability $\phi_0=\phi_1=1/2$.  

A mathematically equivalent definition of the model consists in seeing the dead-leaves construction as a mosaic. With that perspective each grain defines the cell of a tessellation (Fig. \ref{fig:deadleaves}b). The two-phase structure is obtained afterwards by assigning each cell to the pore or solid with probabilities $\phi_0$ or $\phi_1$. A central result of dead-leaves tessellations is the following, which concerns the probability for a given randomly-positioned set $X$ to be included in a single cell of the tessellation \cite{Matheron:1968B,Serra:1982,Jeulin:1993,Jeulin:2000,Lantuejoul:2002}
\begin{equation} \label{eq:deadleaves_prob}
\textrm{Prob} \left\{ X \subseteq \textrm{1 cell} \right\} = \frac{E\left\{ \textrm{Volume}[G \ominus X] \right\}}{E \left\{ \textrm{Volume}[G \oplus X] \right\}}
\end{equation}
where $G \ominus X$ and $G \oplus X$ are the erosion and dilation of the grain by $X$, respectively. The expectation operator $E\left\{ \right\}$ highlights the fact that the grains can have random shape and size.

\begin{figure}
\includegraphics[width=5cm]{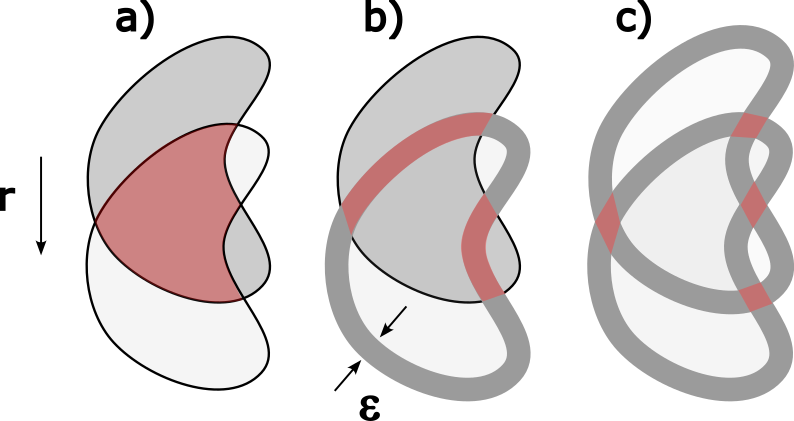}
\caption{Definition of the covariograms $K_v(r)$ (a), $K_{vs}(r)$ (b) and $K_{ss}(r)$ for an arbitrary grain. The grain and the $\epsilon$-thin layer centered on its surface are shown in gray, with translated copies at distance $r$. The intersections relevant to the various covariograms are shown in red.}
\label{fig:covariograms}
\end{figure}

A particular case of Eq. (\ref{eq:deadleaves_prob}) is the probability $P_c(r)$ for two randomly chosen points at distance $r$ from one another to belong to the same cell. This corresponds to a set $X$ consisting in two points at distance $r$ from one another. In that case, Eq. (\ref{eq:deadleaves_prob}) reduces to 
\begin{equation} \label{eq:Pc_deadleave}
P_c(r) = \frac{K_v(r)}{2 K_v(0) - K_v(r)}
\end{equation}
where $K_v(r)$ is the geometrical covariogram of the grain \cite{Serra:1982}, defined as
\begin{equation} \label{eq:K_def}
K_v(r) = E\left\{ \textrm{Volume}[G \ominus X] \right\}
\end{equation}
This is equivalent to defining $K_v(r)$ as the average volume of the intersection of a grain with a replica of it, translated at a distance $r$ away (see Fig. \ref{fig:covariograms}a). The subscript $v$ refers here to the {\it volume} of the grain. It is used here to distinguish that quantity from others that we introduce later, built on the outer surface of the grains, with subscript $s$. 

The two-point pore-pore correlation function of the dead-leaves mosaic is directly obtained the function $P_c(r)$ in Eq. (\ref{eq:Pc_deadleave}), as
\begin{equation} \label{eq:P_to_C}
C_{00}(r) = \phi_0 P_c(r) + \phi_0^2 \left[ 1 - P_c(r) \right]
\end{equation}
This results from the statistical independence of the assignment of each cell to a particular phase. The first term corresponds to two points being in the same cell, which is pore-like with probability $\phi_0$. The second term corresponds to two points in different cells, which are both pore-like with probability $\phi_0^2$. The two-point correlation function of the dead-leaves model is then \cite{Matheron:1968B}
\begin{equation} \label{eq:C11_deadleave}
C_{00}(r) = \phi_0 \phi_1 \frac{K_v(r)}{2 K_v(0) - K_v(r)} + \phi_0^2
\end{equation}
An independent derivation of this equation is provided in Sec. \ref{sec:recursion}, based on a recursive approach. 

Anticipating on the discussion section, various grains are considered in the paper with different covariograms $K_v(r)$ : for monodispersed and polydispersed spheres (Eqs. \ref{eq:Kv_S} and \ref{eq:Kvs_Debye}), disks (Eq. \ref{eq:Kv_D}), and hollow spheres (Eq. \ref{eq:Kv_H}). In all cases, the geometrical covariogram satisfies \cite{Serra:1982}
\begin{equation} \label{eq:Kv_small_r}
K_v(r) \simeq V_G - \frac{A_G}{4} r + \ldots
\end{equation}
for small distances $r$, where $V_G$ and $A_G$ are the average volumes and areas of the grains. The specific surface area of the dead-leaves model is then found to be 
\begin{equation} \label{eq:deadleaves_sd}
s = 2 \phi_0 \phi_1 \frac{A_G}{V_G}
\end{equation}
which results from Eq. (\ref{eq:C00_small_r}), {\it i.e.} from evaluating the derivative of $C_{00}(r)$ in Eq. (\ref{eq:C11_deadleave}) at $r=0$.

\subsection{Alternative approach and surface correlation functions}

\subsubsection{Recursive formalism}
\label{sec:recursion}

We propose here an alternative formalism for the dead-leaves model, which enables one to obtain analytical expressions for the surface correlation functions $F_{0s}(r)$ and $F_{ss}(r)$ as the stable point of an iteration relation. As a validation, we first show that this procedure provides the same expressions for the volume fractions, the correlation functions, and the surface area as in the classical approach. 

At any step $n$ of the dead-leaves construction, the pore indicator function $\mathcal{I}_0^{(n)}(\mathbf{x})$ obeys the following recursion relation
\begin{equation} \label{eq:deadleaves_recursion}
\mathcal{I}_0^{(n+1)}(\mathbf{x}) = \mathcal{I}_0^{(n)}(\mathbf{x}) [1 - \mathcal{G}(\mathbf{x})] + \eta \mathcal{G}(\mathbf{x})
\end{equation}
where $\eta $ is a binary random variable taking values $1$ or $0$ with probabilities $p$ and $1-p$,  and $ \mathcal{G}(\mathbf{x})$ is the indicator function of a yet-unspecified random template structure, which plays the same role as the grains in Fig. \ref{fig:deadleaves}. The interpretation of Eq. (\ref{eq:deadleaves_recursion}) is the following. If $\eta=1$, the values of $I^{(n+1)}_0(\mathbf{x})$ are set to 1 at all points $\mathbf{x}$ in the grains, {\it i.e.} for which $\mathcal{G}(\mathbf{x})=1$. If $\eta=0$, those values of $I^{(n+1)}_0(\mathbf{x})$ are set to 0. The probabilities $p$ and $1-p$ are therefore equivalent to the fraction of pore-like or solid-like grains.

In the dead-leaves construction of Fig. \ref{fig:deadleaves} the template model consists of a dilute dispersion of grains. We keep the discussion more general for now, and let $\mathcal{G}(\mathbf{x})$ be the indicator function of any stationary structure. It has volume fraction
\begin{equation}
\Phi = \langle \mathcal{G}(\mathbf{x}) \rangle
\end{equation}
and two-point correlation function
\begin{equation}
\Gamma(r) = \langle \mathcal{G}(\mathbf{x}) \mathcal{G}(\mathbf{x} + \mathbf{r}) \rangle
\end{equation}
which satisfies $\Gamma(0)=\Phi$, and  $\Gamma(r) \to \Phi^2$ for large $r$. The derivative of $\Gamma(r)$ at the origin is related to the grain's surface area through the usual relation in Eq. (\ref{eq:C00_small_r}).

The pore fraction $\phi_0^{(n)}$ at each step of the construction is calculated through Eq. (\ref{eq:phi_def}) as the average value of the indicator function. From Eq. (\ref{eq:deadleaves_recursion}), the following recursion is obtained
\begin{equation} \label{eq:phi_deadleave}
\phi_0^{(n+1)} = \phi^{(n)}_0 (1 - \Phi) + p \Phi
\end{equation} 
where we took account of the statistical independence of $\eta$ and $\mathcal{G}$ from $\mathcal{I}^{(n)}(\mathbf{x})$, so that the relevant averages of products are factored out as products of averages. The limit of the recursion is obtained as the stable point of Eq. (\ref{eq:phi_deadleave}), namely
\begin{equation}
\phi_0 = p 
\end{equation}
which has an intuitive interpretation. The solid/pore fraction of the asymptotic structure is the equal to the fraction of solid/pore grains used in the construction. 

The two-point correlation function at any step $n$ is calculated as $C^{(n)}_{00}(r) = \langle \mathcal{I}_0^{(n)}(\mathbf{x}) \mathcal{I}_0^{(n)}(\mathbf{x} + \mathbf{r}) \rangle$. From Eq. (\ref{eq:deadleaves_recursion}), the following recursion relation is obtained
\begin{equation} \label{eq:C_deadleaves_recursion}
C_{00}^{(n+1)}(r)= [1 - 2 \Phi + \Gamma(\mathbf{r}) ] C_{00}^{(n)}(r) + 2 p [\Phi - \Gamma(r)] \phi_0^{(n)}  + p \Gamma(r) 
\end{equation}
where we have used the fact that $\eta$ is a binary variable so that $\langle \eta^2 \rangle = \langle \eta \rangle = p$. The limit of $C^{(n)}_{00}(r)$ for $n \to \infty$, is obtained as the stable point of Eq. (\ref{eq:C_deadleaves_recursion}), namely
\begin{equation} \label{eq:C00_infty}
C_{00}(r) = p \frac{\Gamma(r)+ 2 p [\Phi - \Gamma(r)] }{2 \Phi - \Gamma(r)} 
\end{equation}
where we used the fact that asymptotic value of $\phi^{(n)}_0$ for large $n$ is $p$. From the general relation $\Gamma(0)= \Phi$, one checks that this expression has the expected limit for small $r$, namely $C_{00}(0) = p = \phi_0$. Interestingly, it does not generally have the correct value for large $r$. Using the general limit $ \Gamma(r \to \infty) = \Phi^2$, one finds indeed
\begin{equation} \label{eq:C_00}
\lim_{r \to \infty} C_{00}(r) = p \frac{\Phi + 2 p (1-\Phi)}{\Phi + 2 (1-\Phi)}
\end{equation}
which is equal to $p^2=\phi_0^2$ only in the limit of a grain template with vanishingly small volume fraction $\Phi$. 

In order for the limit in Eq. (\ref{eq:C_00}) to be equal to $p^2 $, one has to use a template with vanishingly small volume fraction $\Phi \ll 1$. One possibility consists in using a dilute dispersion of actual grains, as in the original dead-leaves model. In that case, the volume fraction of the template is
\begin{equation} \label{eq:Phi}
\Phi = \theta V_G + \mathcal{O}(\theta^2)
\end{equation}
where $V_G$ is the average volume of the grains, and $\theta$ is their density assumed to be vanishingly small. The correlation function of the template is
\begin{equation} \label{eq:Gamma_grain}
\Gamma(r) = \theta K_v(r) + \mathcal{O}(\theta^2)
\end{equation}
where $K_v(r)$ is the geometrical covariogram of the grain, defined as in Eq. (\ref{eq:K_def}). Equations (\ref{eq:Phi}) and (\ref{eq:Gamma_grain}) can be understood as the small-$\theta$ limit of a Boolean model of interpenetrating grains \cite{Matheron:1967,Serra:1982,Torquato:2002}.  Using these expressions for $\Phi$ and $\Gamma(r)$, the asymptotic correlation function in Eq. (\ref{eq:C00_infty}) becomes
\begin{equation}
C_{00}(r) = p (1-p) \frac{K_v(r)}{2K_v(0) - K_v(r)} + p^2 
\end{equation}
which is identical to the classical result in Eq. (\ref{eq:C11_deadleave}), as it should.

The same iterative approach can be used to characterize the surface of the dead-leaves model. For that purpose, we define $\partial \mathcal{G}_\epsilon(\mathbf{x})$ to be the indicator function of the border of the grains, with thickness $\epsilon$. For dilute grains, the volume fraction of the grains' border $\partial \Phi_{\epsilon} = \langle \partial \mathcal{G}_\epsilon(\mathbf{x})\rangle $ is calculated as
\begin{equation} \label{eq:Phi_B}
\partial \Phi_\epsilon = \theta A_G \times \epsilon + \mathcal{O}(\theta^2)
\end{equation}
where $A_G$ is the average area of the individual grains. The terms of order $\theta^2$ and higher account for grain overlapping, and are negligibly small in the dilute regime. With these notations and assumptions, the indicator function of the surface layer at step $n$ satisfies the following recursion equation
\begin{align} \label{eq:deadleaves_recursion_S}
\mathcal{S}_{\epsilon}^{(n+1)}(\mathbf{x}) & = \mathcal{S}_{\epsilon}^{(n)}(\mathbf{x}) [1- \mathcal{G}(\mathbf{x}) ]  \cr
&+ \partial \mathcal{G}_\epsilon (\mathbf{x}) \Big\{  \eta  [1-\mathcal{I}_0^{(n)}(\mathbf{x}) ] + (1-\eta) \mathcal{I}_0^{(n)}(\mathbf{x})  \Big\}
\end{align}
The first term on the right-hand side sets $\mathcal{S}_\epsilon^{(n+1)}(\mathbf{x})$ to zero inside the newly-positioned grain $\mathcal{G}(\mathbf{x})$. The second term sets its value to one wherever a new surface is created: depending on the value of $\eta$, this is either in the pre-existing solid (where $1-\mathcal{I}_0^{(n)}(\mathbf{x}) = 1$) or in the pore phase (where $\mathcal{I}_0^{(n)}(\mathbf{x}) = 1$). Note that Eq. (\ref{eq:deadleaves_recursion_S}) assumes that the $\epsilon$-thin layer in $\partial \mathcal{G}_\epsilon (\mathbf{x})$ is entirely on the inner side of the surface. As we discuss later, this is an inconsequential yet mathematically convenient choice. 

The surface area of the model $s^{(n)}$ at construction step $n$, is calculated from the volume fraction of the surface layer as
\begin{equation}
\phi_{S_\epsilon}^{(n)} = s^{(n)} \times \epsilon
\end{equation}
where $\phi_{S_\epsilon} ^{(n)}= \langle \mathcal{S}_{\epsilon}^{(n)}(\mathbf{x}) \rangle$. Evaluating the average value of Eq. (\ref{eq:deadleaves_recursion_S}), provides the following recursion for the volume fraction of the surface layer
\begin{equation}
\phi_{S_\epsilon}^{(n+1)} =  [1- \Phi ] \phi_{S_\epsilon}^{(n)} + (1-2p) \partial  \Phi_\epsilon \phi_0^{(n)} + p \partial  \Phi_\epsilon
\end{equation}
Evaluating the limit for $n \to \infty$ as the stable point of the recursion, provides the following asymptotic value 
\begin{equation}
\phi_{S_\epsilon} = \frac{\partial  \Phi_\epsilon}{\Phi} \times 2 p (1-p)
\end{equation}
where we have used the fact that the limit of $\phi_0^{(n)}$ for large $n$ is equal to $p$. Based on Eqs. (\ref{eq:Phi}) and (\ref{eq:Phi_B}), this is identical to the classical expression of the surface area of the dead-leaves model in Eq. (\ref{eq:deadleaves_sd}).

\subsubsection{Pore-surface correlation function}

The pore-surface correlation function can be calculated starting from the two-point correlation function $C_{0S_\epsilon}(r)$ between the pore space and the surface layer $\mathcal{S}_\epsilon(\mathbf{x})$, as defined in Eq. (\ref{eq:C0S_def}). Combining the recursion relations for $\mathcal{I}(\mathbf{x})$ and $\mathcal{S}_{\epsilon}(\mathbf{x})$ - from Eq. (\ref{eq:deadleaves_recursion}) and (\ref{eq:deadleaves_recursion_S}) , respectively - one obtains the following recursion relation for $C_{0S_\epsilon}(r)$
\begin{align} \label{eq:C0S_recursion}
C_{0S_\epsilon}^{(n+1)}(r) &= \left[1 - 2 \Phi + \Gamma(\mathbf{r}) \right] C_{0S_\epsilon}^{(n)}(r) 
+ (1-2p) [\partial  \Phi_\epsilon - \partial  \Gamma_\epsilon (r) ] C_{00}^{(n)}(r) \cr
&+ p [ \partial  \Phi_\epsilon - 2 \partial  \Gamma_\epsilon (r) ] \phi_0^{(n)}  
+ p \left[ \Phi - \Gamma(r) \right] \phi_{S_\epsilon}^{(n) } \cr
&+ p \partial  \Gamma_\epsilon (r)
\end{align}
where
\begin{equation}
\partial  \Gamma_\epsilon(r)  = \langle \mathcal{G}(\mathbf{x})  \partial  \mathcal{G}_\epsilon(\mathbf{x}+\mathbf{r}) \rangle 
\end{equation}
is the cross-correlation function between the grains and their $\epsilon$-thin surface layer. 

In the case of dilute grains, relevant to the dead-leaves model, one has the following limit
\begin{equation} \label{eq:lim_Kvs}
\lim_{\epsilon \to 0} \frac{\partial \Gamma_\epsilon(r) }{\epsilon} = \theta K_{vs}(r) + \mathcal{O}(\theta^2)
\end{equation}
where $K_{vs}(r)$ is the cross-correlation between the grain volume and its surface (see Fig. \ref{fig:covariograms}b). The terms of order $\theta^2$ or higher correspond to one point inside a grain, and the other on the boundary of another grain. These terms are negligible in the limit of vanishingly small $\theta$. 

We consider various types of grains in the rest of the paper, with different volume-surface covariograms $K_{vs}(r)$ : for monodispersed and polydispersed spheres (Eqs. \ref{eq:Kvs_S} and \ref{eq:Kvs_Debye}), disks (Eq. \ref{eq:Kvs_D}), and hollow spheres (Eq. \ref{eq:Kvs_H}). In all cases, the volume-surface covariogram $K_{vs}(r)$ has the following asymptotic form for small distances
\begin{equation} \label{eq:Kvs_small_r}
K_{vs}(r) \simeq \frac{A_G}{2} \left( 1 - \frac{H_G}{2}  r + \ldots \right)
\end{equation}
which is formally identical to the asymptotic form of the pore-surface correlation function in Eq. (\ref{eq:F0s_small_r}), where $H_G$ is here the average mean curvature of the grain. 

The asymptotic limit of $C_{0S_\epsilon}^{(n)}(r)$ for $n \to \infty$, is calculated as the stable point of Eq. (\ref{eq:C0S_recursion}), from which the pore-surface correlation function in calculated as $\lim_{\epsilon \to 0} C_{0S\epsilon}(r)/\epsilon$. The result can be written as
\begin{align} \label{eq:deadleaves_F0s}
F_{0s}(r) &= s \left[ \phi_0 + \frac{1}{2} (1 - 2 \phi_0) \frac{C_{00}(r)-\phi_0^2}{\phi_0 \phi_1} \frac{K_v(0)}{2 K_v(0) - K_v(r) }\right] \cr
&- (1-2\phi_0) [ C_{00}(r) - \phi_0 ] \frac{K_{vs}(r) }{2 K_v(0) - K_v(r) }
\end{align}
where we have used the notation $K_v(0)$ for the volume of the grain $V_G$. This expression of the pore-surface correlation function $F_{0s}(r)$ satisfies the two limits $F_{0s}(r \to 0) = s/2$ and $F_{0s}(r \to \infty) = \phi_0 s$. Moreover, in the particular case of $\phi_1 = \phi_0 = 1/2$, it also reduces to $F_{0s}(r) = s/2$ for all $r$. This is the expected mathematical consequence of the phase-exchange symmetry of the dead-leaves model.

As a technical comment, it is worth mentioning that Eq. (\ref{eq:lim_Kvs}) strictly holds only for $r>0$. This results from the layer $\partial \mathcal{G}_\epsilon(\mathbf{x})$ being defined on the inner side of the surface of $\mathcal{G}(\mathbf{x})$, which is implicit in Eq. (\ref{eq:deadleaves_recursion_S}). For $r=0$ the limit in the left-hand side of Eq. (\ref{eq:lim_Kvs}) is therefore equal to $\theta A_G$. This contrasts with the volume-surface covariogram $K_{vs}(r)$ in the right-hand side, which is defined with the layer centered on the surface and takes the value $K_{vs}(0)=A_G/2$. These differences are irrelevant for any finite $r$. The expression of the pore-surface correlation function in Eq. (\ref{eq:deadleaves_F0s}) is immune to that discontinuity at $r=0$ because $K_{vs}(r)$ is multiplied by $C_{00}(r) - \phi_1$, which vanishes for $r=0$. 

\subsubsection{Surface-surface correlation function}

The surface-surface correlation function can be calculated starting from the two-point correlation function of the surface layer, $C_{S_\epsilon S_\epsilon}(r)$ defined in Eq. (\ref{eq:CSS_def}). From the recursion relation for $\mathcal{S}_{\epsilon}(\mathbf{x})$ in Eq. (\ref{eq:deadleaves_recursion_S}), one obtains the following recursion relation
\begin{align} \label{eq:CSS_recursion}
C_{S_\epsilon S_\epsilon}^{(n+1)}(r) &= \left[1 - 2 \Phi + \Gamma(\mathbf{r}) \right] C_{S_\epsilon S_\epsilon}^{(n)}(r) 
+  \partial^2  \Gamma_\epsilon (r)  C_{00}^{(n)}(r)
+ 2 (1-2p) \left[ \partial \Phi_\epsilon - \partial \Gamma_\epsilon (r) \right] C_{0S_\epsilon}^{(n) } \cr
& + 2 p \left[ \partial \Phi_\epsilon - \partial \Gamma_\epsilon (r) \right] \phi_{S_\epsilon}^{(n) } 
- 2 p  \partial^2  \Gamma_\epsilon (r) \phi_0^{(n)} 
+ p \partial^2  \Gamma_\epsilon (r)
\end{align}
where
\begin{equation}
\partial^2  \Gamma_\epsilon(r)  = \langle \partial \mathcal{G}(\mathbf{x})  \partial  \mathcal{G}_\epsilon(\mathbf{x}+\mathbf{r}) \rangle 
\end{equation}
is the self-correlation function of the grains $\epsilon$-thin surface layer. 

In the case of dilute grains, relevant to the dead-leaves model, one has the following limit
\begin{equation}
\lim_{\epsilon \to 0} \frac{\partial^2 \Gamma_\epsilon(r) }{\epsilon^2} = \theta K_{ss}(r) + \mathcal{O}(\theta^2)
\end{equation}
where $K_{ss}(r)$ is the self-correlation of the surface of individual grains (see Fig. \ref{fig:covariograms}c). The surface-surface covariogram satisfies the general relation
\begin{equation} \label{eq:Kss_small_r}
K_{ss}(r) \simeq \frac{A_G}{2r}
\end{equation}
for small values or $r$. This is the equivalent for a grain of the asymptotic form of the surface correlation function in Eq. (\ref{eq:Fss_small_r}). The various expressions of $K_{ss}(r)$ considered in the rest of the paper include mono-and poly-dispersed spheres in Eqs. (\ref{eq:Kss_S}) and (\ref{eq:Kvs_Debye}), disks in (Eq. \ref{eq:Kss_D}), and hollow spheres in Eq. (\ref{eq:Kss_H}). 

The limit of $C_{S_\epsilon S_\epsilon}^{(n)}(r)$ for $n \to \infty$ is the stable point of Eq. (\ref{eq:CSS_recursion}). The surface-surface correlation function is then obtained as $\lim_{\epsilon \to 0} C_{S_\epsilon S_\epsilon}(r)/\epsilon^2$. The result can be written as
\begin{align} \label{eq:deadleaves_Fss}
F_{ss}(r) &= \left[ C_{00}(r) - \phi_0^2 + \phi_1 \phi_0 \right] \frac{K_{ss}(r)}{2K_v(0) - K_v(r)}  \cr
&+2\left[ \phi_0 s + (1- 2 \phi_0) F_{0s}(r) \right] \frac{2 K_{vs}(0)-K_{vs}(r)}{2K_v(0) - K_v(r)}
\end{align}
where we have used the notation $2K_{vs}(0)$ for the volume of the grain $A_G$. This expression of the surface-surface correlation function $F_{ss}(r)$ converges to $s^2$ for large values of $r$, as it should. It also satisfies the general asymptotic Eq. (\ref{eq:Fss_small_r}) for small $r$. Moreover, one can also check that it is fully symmetric with respect to exchanging the two phases $1$ and $0$. 

\begin{figure}
\includegraphics[width=12cm]{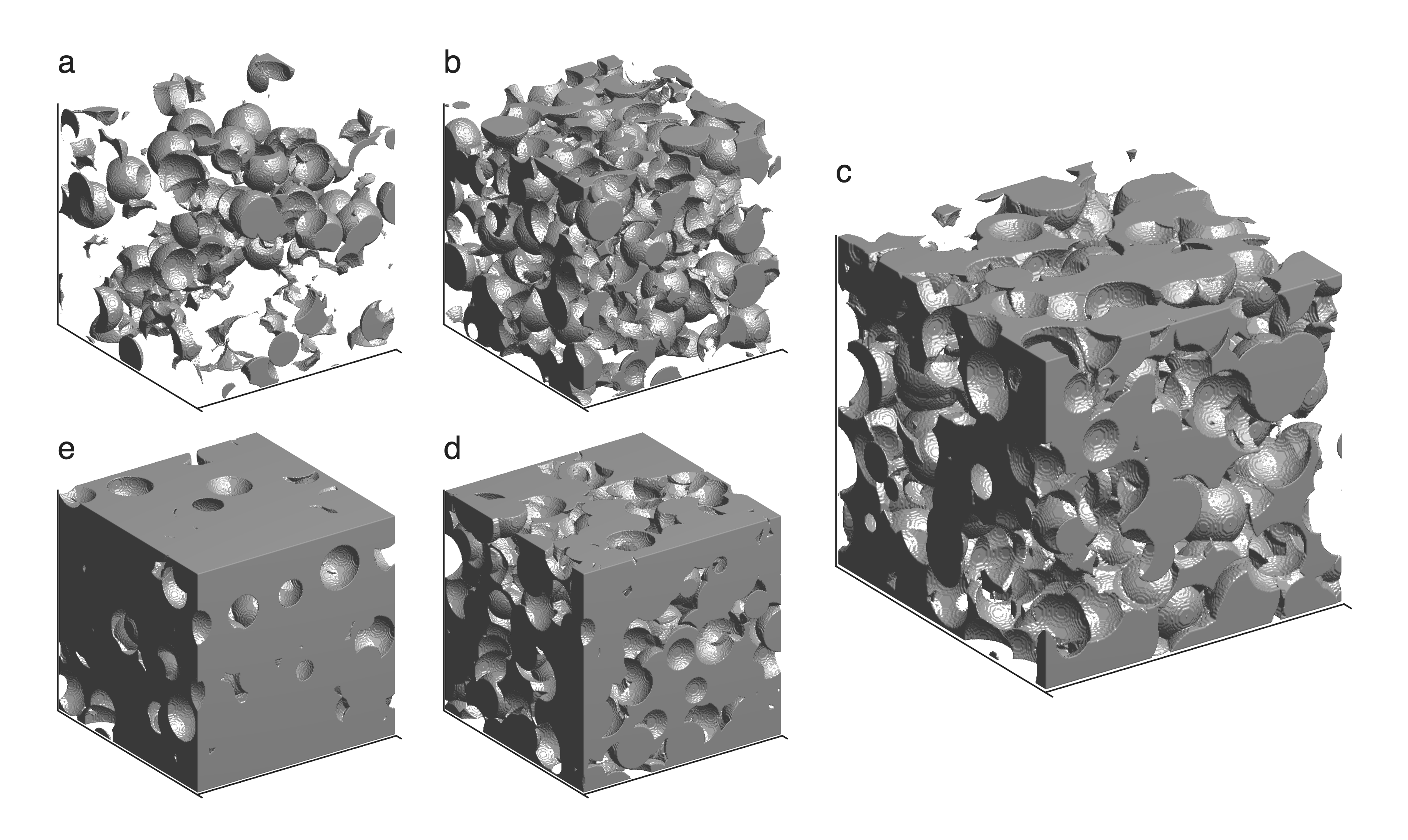}
\caption{Realizations of dead-leaves models from monodispersed spherical grains, with decreasing porosities $\phi_0=0.9$ (a), $\phi_0=0.7$ (b), $\phi_0=0.5$ (c),  $\phi_0=0.3$ (d) and $\phi_0=0.1$ (e). Note the phase-inversion symmetry of a-e and b-d.}
\label{fig:realizations}
\end{figure}

\section{Discussion}

The main results of the paper are Eqs. (\ref{eq:deadleaves_F0s}) and (\ref{eq:deadleaves_Fss}), which are exact analytical expressions for the pore-surface and the surface-surface correlation functions of the dead-leaves model. These results are valid for arbitrary grains, in arbitrary dimension. 

\begin{figure}
\includegraphics[width=7cm]{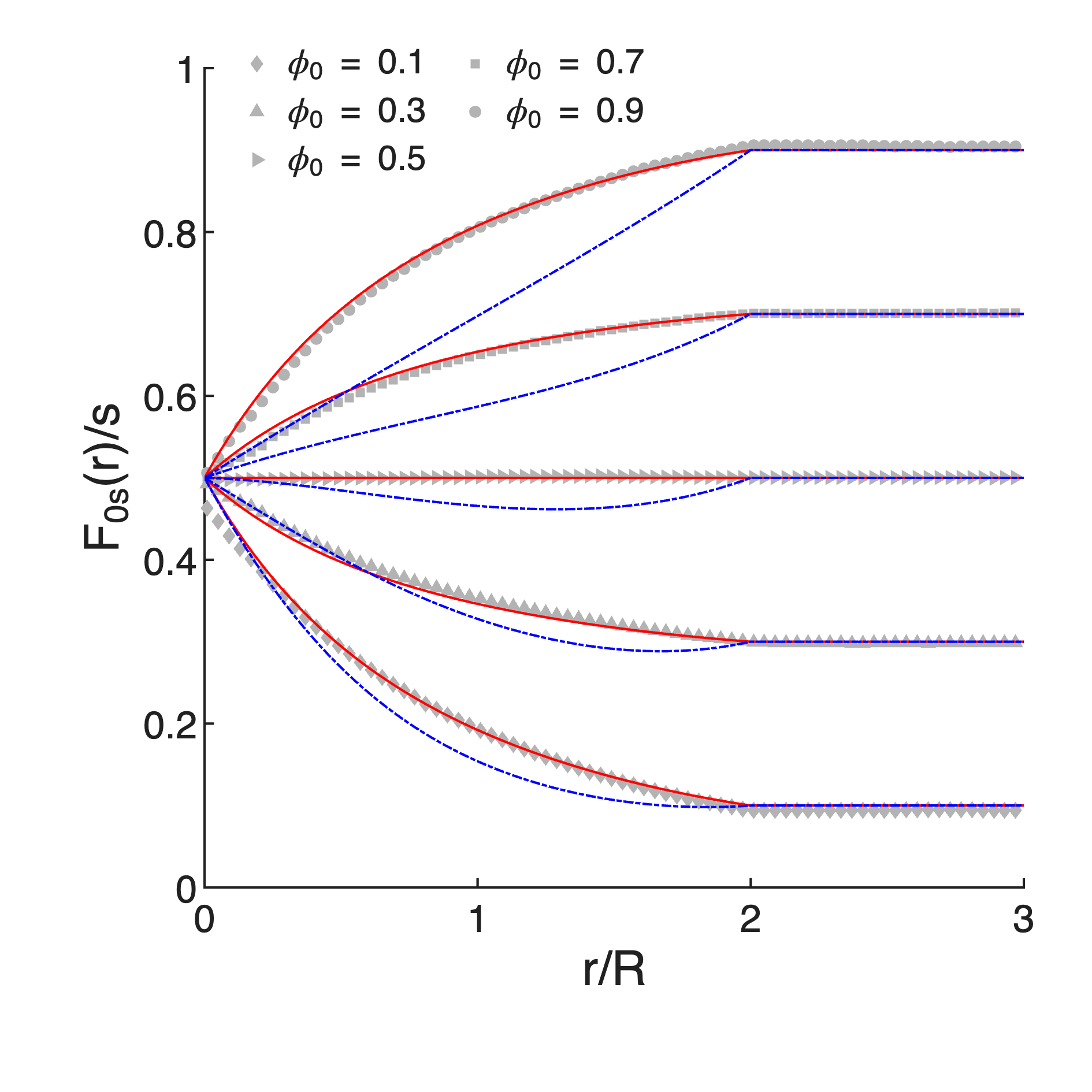}
\caption{Pore-surface correlation functions $F_{0s}(r)$ of the dead-leaves model, with spherical grains with radius $R$ and various porosities $\phi_0$. The symbols are measured on numerical realizations and the solid red lines are calculated from Eq. (\ref{eq:deadleaves_F0s}). For comparison, the values for Boolean models with identical porosities and grain size are plotted as dashed blue lines.}
\label{fig:deadleaves_F0s}
\end{figure}

\begin{figure}
\includegraphics[width=7cm]{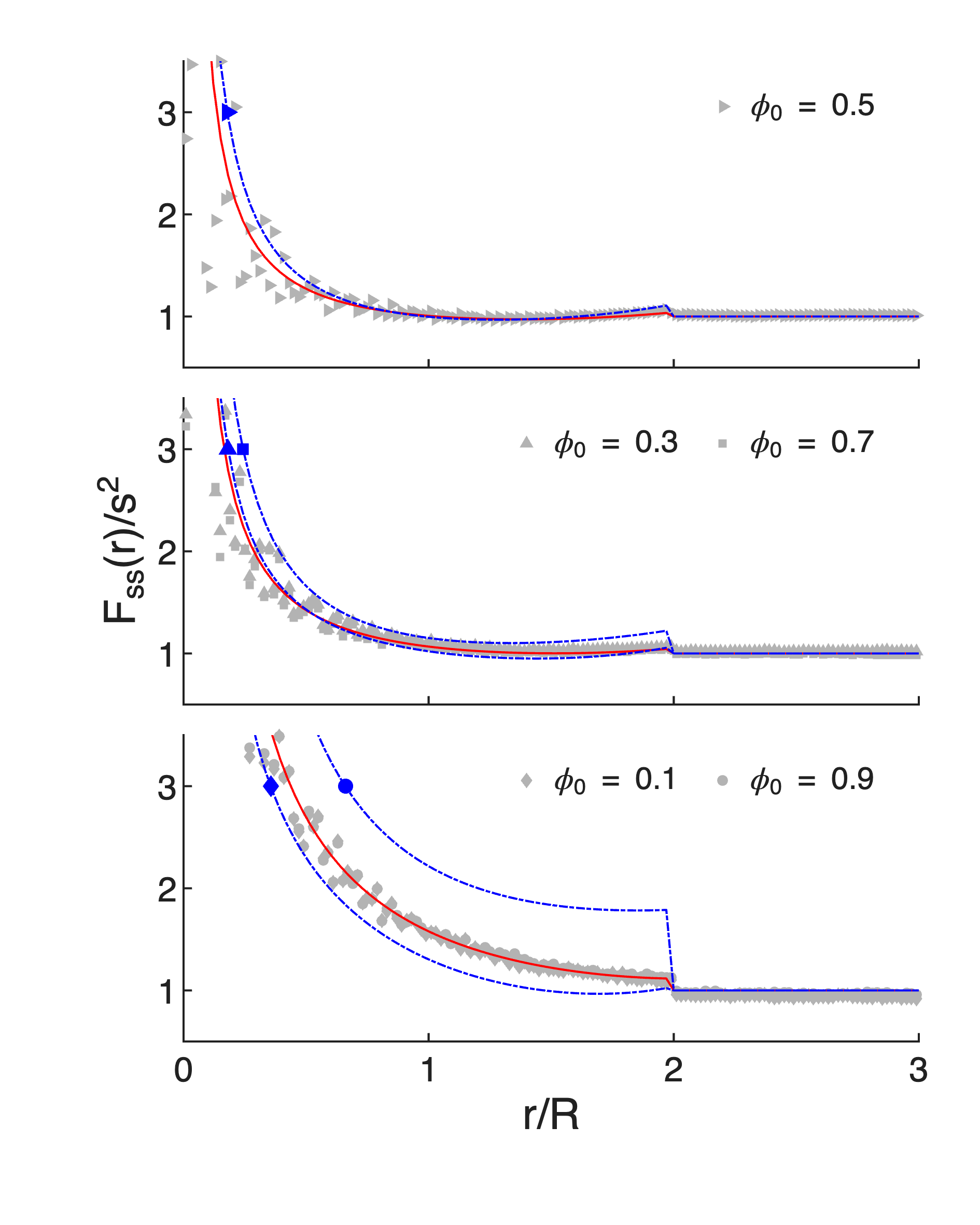}
\caption{Surface-surface correlation functions $F_{ss}(r)$ of the dead-leaves model, with spherical grains with radius $R$ and various porosities $\phi_0$. The symbols are measured on numerical realizations and the solid red lines are calculated from Eq. (\ref{eq:deadleaves_Fss}). For comparison, the values for Boolean models with identical porosities and grain size are plotted as dashed blue lines.}
\label{fig:deadleaves_Fss}
\end{figure}

In the particular case of a dead-leaves model built from monodispersed spherical grains, the relevant values of the volume and surface covariograms are the following 
\begin{equation} \label{eq:Kv_S}
K_v^{(S)}(r) = \frac{4 \pi R^3}{3} \left( 1 - \frac{r}{2R} \right)^2 \left(1+ \frac{r}{4R} \right) \Theta(2R-r)
\end{equation}
\begin{equation} \label{eq:Kvs_S}
K_{vs}^{(S)}(r) = 2 \pi R^2 \left(1 -\frac{r}{2R} \right) \Theta(2R-r)
\end{equation}
\begin{equation} \label{eq:Kss_S}
K_{ss}^{(S)}(r) = \frac{2 \pi R^2}{r} \Theta(2R-r)
\end{equation} 
where $\Theta(x)$ is the step function, equal to 1 for $x \ge 0$ and to 0 otherwise. Equation (\ref{eq:Kv_S}) is the classical intersection volume of two spheres with radius $R$ (see Fig. \ref{fig:covariograms}a). Equation (\ref{eq:Kvs_S}) is the intersection volume of a sphere with its $\epsilon$-thin surface layer, normalized by $\epsilon$ (see Fig. \ref{fig:covariograms}b). Equation (\ref{eq:Kss_S}) is the intersection volume the surface layer, normalized by $\epsilon^2$ (see Fig. \ref{fig:covariograms}c). The resulting pore-surface and surface-surface correlation functions $F_{0s}(r)$ and $F_{ss}(r)$ are plotted in Fig. \ref{fig:deadleaves_F0s} and in Fig. \ref{fig:deadleaves_Fss} for various porosities $\phi_0$. 

Realizations of the dead-leaves model with spherical grains are shown in Fig. \ref{fig:deadleaves_F0s}, and numerically-evaluated values of $F_{0s}(r)$ and $F_{ss}(r)$ are also plotted in Figures \ref{fig:deadleaves_F0s} and \ref{fig:deadleaves_Fss}. These values result from 40 independent realizations averaged for each condition, with size $512^3$ and grain radius $R=50$, on which the surface correlation functions were measured using the Matlab code developed in Ref.~\cite{Ma:2018}. Because absolute values of the surface areas are poorly captured by the numerical methods \cite{Samarin:2023}, the plots are normalized to the surface areas $s$, estimated from the numerical values of $F_{0s}(r)$ and $F_{ss}(r)$ at large $r$. 

The functions in Figs. \ref{fig:deadleaves_F0s} and \ref{fig:deadleaves_Fss}  testify to the phase-inversion symmetry of the dead-leaves model. In the case of the surface-surface function $F_{ss}(r)$, the values are identical for complementary values of $\phi_0$. In the case of the pore-surface function $F_{0s}(r)$, the values for complementary $\phi_0$'s are vertically symmetric around the constant value $F_{0s}(r) = s/2$, which is exactly satisfied for $\phi_0=1/2$. In addition to phase-inversion symmetry, it is also noteworthy that the pore-surface correlation functions $F_{0s}(r)$ in Fig. \ref{fig:deadleaves_F0s} all have the same shape. This is evidenced by rewriting Eq. (\ref{eq:deadleaves_F0s}) in the following normalized form
\begin{equation} \label{eq:F0s_mosaic_DL}
\frac{F_{0s}(r) - \phi_0 s}{(1/2-\phi_0) s} = \left[\frac{K_v(r)}{K_v(0)} + \frac{K_{vs}(r)}{K_{vs}(0)} - \frac{K_v(r)}{K_v(0)} \frac{K_{vs}(r)}{K_{vs}(0)}\right] \times \left[ 2 - \frac{K_v(r)}{K_v(0)} \right]^{-2}
\end{equation}
where the left-hand side takes values between 1 for $r=0$ and 0 for $r \to \infty$. The interesting point is that the right-hand side depends only on the grain characteristics, but not on the porosity $\phi_0$. We show in Appendix \ref{sec:mosaic} that this very specific type of dependence on $\phi_0$ holds for any mosaic structure. The right-hand side of Eq. (\ref{eq:F0s_mosaic_DL}) is related to the probability for two points at distance $r$ from one another to be in a cell of the underlying tessellation, and on the boundary of the same cell. 

For comparison, the surface correlation functions of the Boolean model of spheres are also plotted in Figs. \ref{fig:deadleaves_F0s} and \ref{fig:deadleaves_Fss}. Using the same notations as in the rest of the paper, they are written as \cite{Doi:1976,Torquato:2002,Ma:2018} 
\begin{align}  \label{eq:Boolean_F0s}
F_{0s}(r) = \frac{s}{\phi_0} \left[ 1 - \frac{K_{vs}(r)}{2 K_{vs}(0)}\right] C_{00}(r)
\end{align}
and
\begin{align}  \label{eq:Boolean_Fss}
F_{ss}(r) = \left\{ \left(\frac{s}{\phi_0}\right)^2 \left[ 1 - \frac{K_{vs}(r)}{2 K_{vs}(0)}\right]^2 + \theta K_{ss}(r) \right\} C_{00}(r)
\end{align}
where $\theta$ is density of the grains, related to the porosity via
\begin{equation} \label{eq:Boolean_phi0}
\phi_0 = \exp\left[ -\theta K_v(0) \right]
\end{equation}
and the two-point correlation function of the pores of the Boolean model is 
\begin{equation} \label{eq:Boolean_C00}
C_{00}(r) = \phi_0^2 \exp\left[ \theta K_v(r) \right]
\end{equation}
The two last equations are valid for arbitrary grains in any dimension \cite{Matheron:1967,Serra:1982,Lantuejoul:2002,Torquato:2002}. Although Eqs (\ref{eq:Boolean_F0s}) and (\ref{eq:Boolean_Fss}) have been proven only for monodispersed spheres \cite{Doi:1976} - {\it i.e.} for the specific covariograms in Eqs. (\ref{eq:Kv_S}), (\ref{eq:Kvs_S}) and (\ref{eq:Kss_S}) - we argue in Appendix \ref{sec:Boolean} that they are valid for arbitrary grains if they are expressed in terms of $K_v(r)$, $K_{vs}(r)$ and $K_{ss}(r)$.

When looking at the pore-surface correlation functions in Fig. \ref{fig:deadleaves_F0s}, a striking difference between the dead-leaves and the Boolean models is the different slope of at the origin, pointing at different {\it apparent} curvatures $\bar H$. The word  ``apparent'' is a caveat that we introduced when discussing Eq. (\ref{eq:F0s_small_r}), to account for the possible presence of singularities on the surface. Only for smooth surfaces, that is without singularities, can the slope of $F_{0s}(r)$ for $ r\to 0$ be unambiguously interpreted as the average mean curvature \cite{Teubner:1990,Ma:2018}. Singularities are present in the surfaces of both the Boolean and dead-leaves models, where the grains overlap (see Fig. \ref{fig:realizations}). 

From Eq. (\ref{eq:F0s_small_r}), as well as Eqs. (\ref{eq:Kv_small_r}) and (\ref{eq:Kvs_small_r}) which hold for arbitrary three-dimensional grains, the following value is obtained for the dead-leaves model
\begin{equation} \label{eq:deadleaves_H}
\bar H = (1-2 \phi_0) \frac{A_G}{V_G}
\end{equation}
where the $1-2 \phi_0$ dependence results from the fact that the dead-leaves structure is a mosaic (see Eq. \ref{eq:F0s_mosaic_DL}). When applying the same procedure to the Boolean model, one finds
\begin{equation} \label{eq:Boolean_H}
\bar H = - H_G - \frac{1}{2} \frac{A_G}{V_G} \ \ln(\phi_0) 
\end{equation}
It has to be stressed that Eq. (\ref{eq:deadleaves_H}) holds for arbitrary grains, so that the mean curvature $H_G$, area $A_G$ and volume $V_G$ are independent characteristics of the grains. In that respect, it is striking that the curvature of the dead-leaves structure does not depend on the curvature of the individual grains $H_G$. This is a consequence of the strong overlapping of the grains, which is inherent to the dead-leaves construction for any porosity (see Fig. \ref{fig:deadleaves}c). The situation is different for the Boolean model, because no overlapping happens in the dilute regime. The average mean curvature of the structure has therefore to converge to that of the individual grains in the limit of $\phi_0 \to 1$, as is indeed the case for Eq. (\ref{eq:Boolean_H}).

\begin{figure}
\includegraphics[width=8cm]{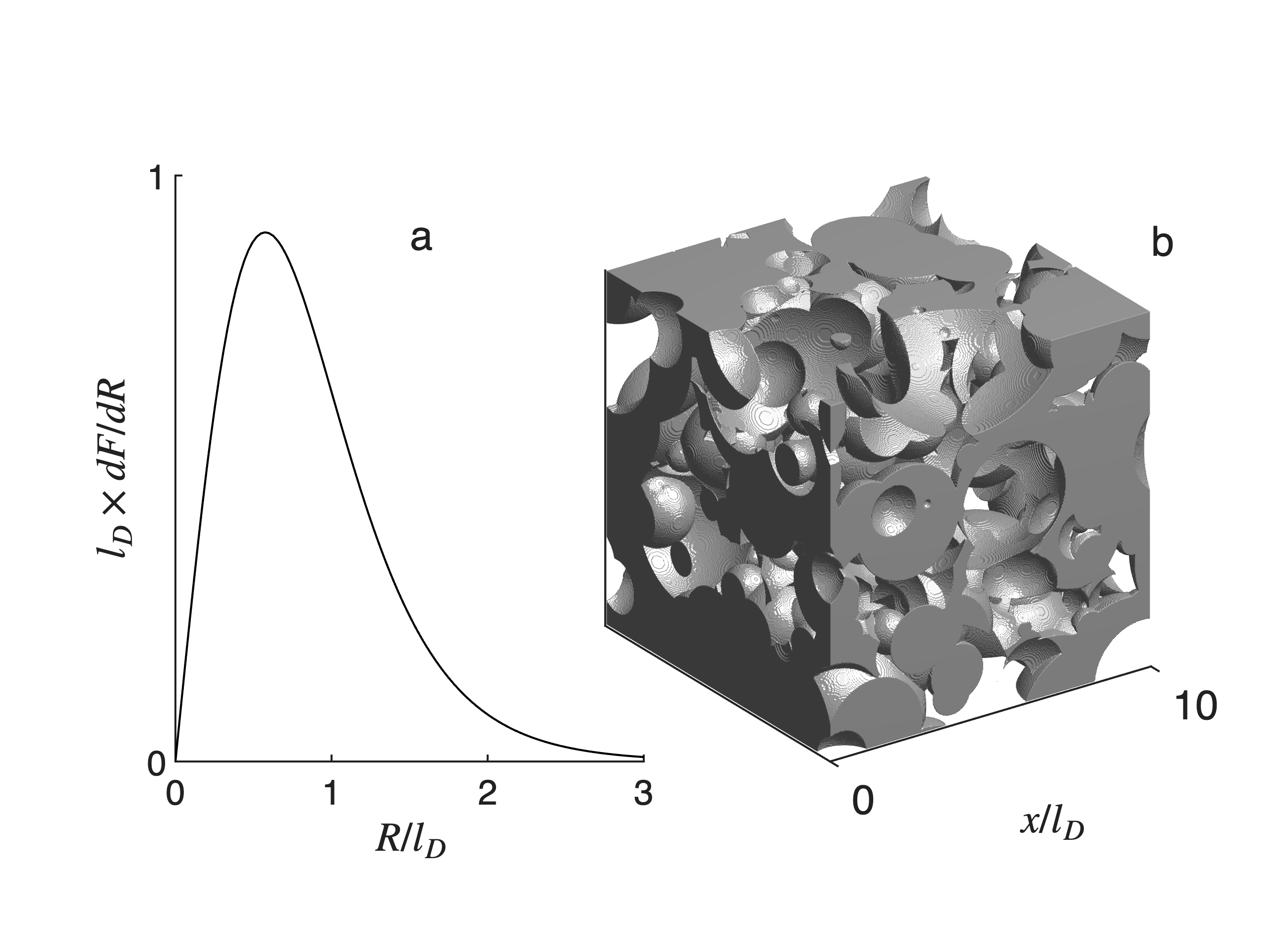}
\caption{Debye random medium realized as a dead-leaves structure: (a) spherical grain size distribution yielding exponential two-point correlation function, and (b) particular realization for $\phi_0=0.5$.}
\label{fig:Debye_demo}
\end{figure}

That the dead-leaves model can be defined for arbitrary grains, offers the possibility of exploring the properties of structures with targeted two-point correlation functions. An interesting case to consider is the so-called Debye random medium \cite{Ciccariello:1983,Levitz:1992,Yeong:1998,Ma:2020,Skolnick:2021}, which is defined as a random structure having exponential two-point correlation function, namely 
\begin{equation} \label{eq:Debye:C00}
C_{00}(r) = \phi_0 \phi_1 \exp\left[ -r/l_D \right] + \phi_0^2
\end{equation}
where $l_D$ is the only parameter of the model, having the meaning of a characteristic length. From the general expression of the two-point correlation function in Eq. (\ref{eq:C11_deadleave}) a dead-leaves structure with exponential two-point correlation function can in principle be obtained for any porosity, by choosing a grain with geometrical covariogram of the type
\begin{equation} \label{eq:Kv_Debye}
\frac{K_v^{(D)}(r)}{K_v^{(D)}(0)} = \frac{2}{1 + \exp[r/l_D]}
\end{equation}
It was shown that this covariogram can indeed be realized as polydispersed spheres with the following cumulative radius distribution \cite{Emery:2010,Gommes:2018}
\begin{equation} \label{eq:FR_3}
F(R) = 1 - 8 \frac{\exp[2R/l_D] (\exp[2R/l_D] -1)}{(2R/l_D) (\exp[2R/l_D] +1)^3}
\end{equation}
that is, $F(R)$ is the probability for a grain to have radius smaller than $R$. The probability density function $dF/dR$ and a particular realization of the structure are shown in Fig. \ref{fig:Debye_demo}.

The area and volume of the so-defined ``Debye'' grains, obtained from the second and third moments of $F(R)$, are
\begin{equation}
A_G^{(D)} = 4 \pi l_D^2 \quad \textrm{and} \quad V_G^{(D)} = 2 \pi l_D^3
\end{equation}
and the resulting apparent mean curvature from Eq. (\ref{eq:deadleaves_H}) is 
\begin{equation} \label{eq:Debye_H}
\bar H = (1/2 - \phi_0) \frac{4}{l_D}
\end{equation}
The volume-surface and surface-surface covariograms of the Debye grains are obtained by the following convolutions
\begin{equation} \label{eq:Kvs_Debye}
K_{vs}^{(D)}(r) = \int_0^1 K_{vs}^{(S)}(r) \ dF(R) \quad \textrm{and} \quad K_{ss}^{(D)}(r) = \int_0^1 K_{ss}^{(S)}(r) \ dF(R)
\end{equation}
where $K_{vs}^{(S)}(r)$ and $K_{ss}^{(S)}(r)$ are the covariograms of the spheres given in Eqs. (\ref{eq:Kvs_S}) and (\ref{eq:Kss_S}). 

\begin{figure}
\includegraphics[width=7cm]{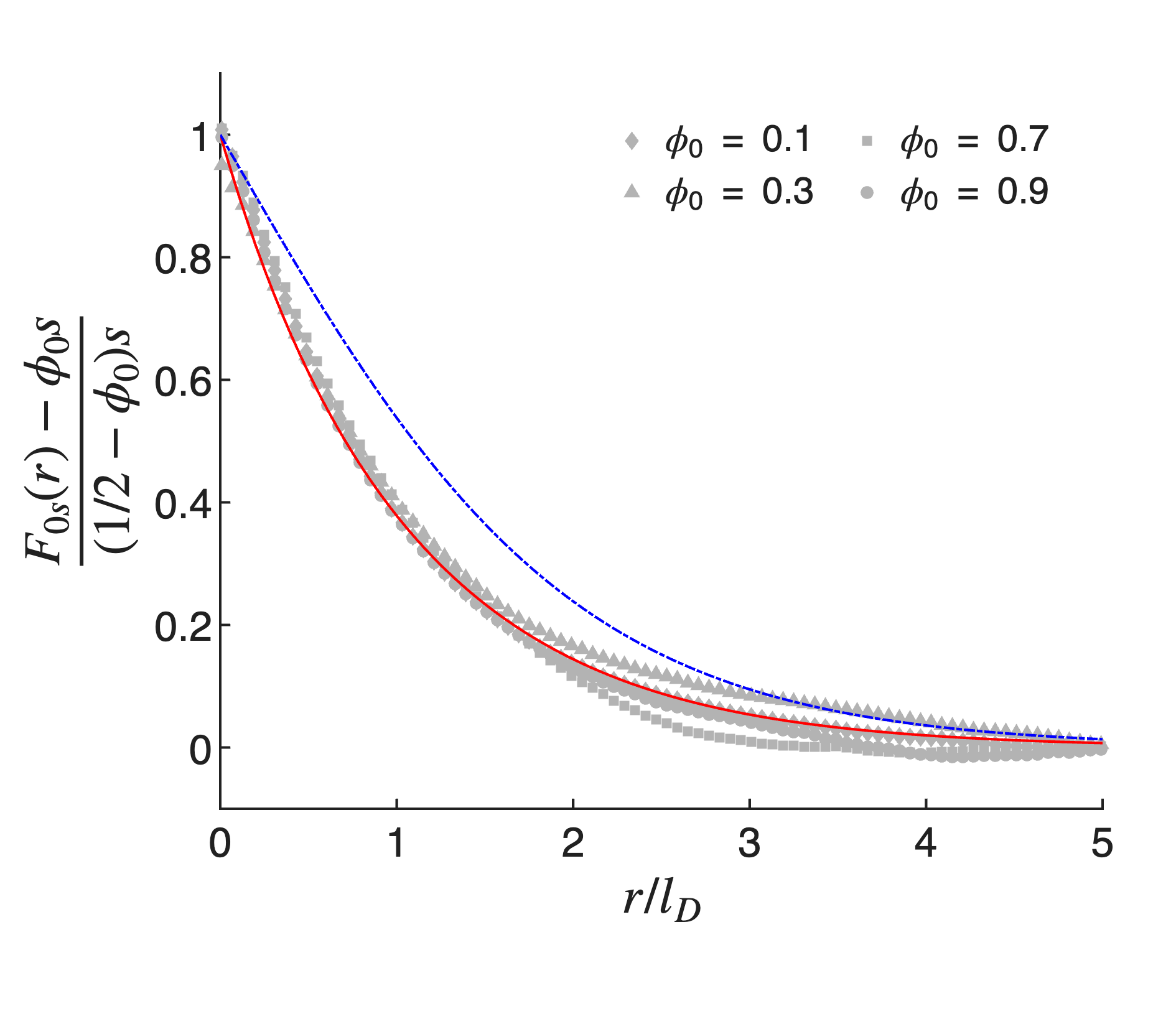}
\caption{Pore-surface correlation functions $F_{0s}(r)$ of Debye random media obtained as a dead-leaves model (solid red line) or via simulated annealing (dashed blue). The symbols are measured from realizations of the dead-leaves model.}
\label{fig:Debye_F0s}
\end{figure}

The pore-surface correlation function of the dead-leaves model with exponential covariance is plotted in Fig. \ref{fig:Debye_F0s}. For comparison, the values of $F_{0s}(r)$ obtained in Ref.~\cite{Ma:2020} via simulated-annealing reconstruction of Debye random media are also plotted in Fig. \ref{fig:Debye_F0s}. The semi-empirical expression of $F_{0s}(r)$ proposed in that paper, can be put in the following reduced form 
\begin{equation} \label{eq:Debye_F0s}
\frac{F^{(D)}_{0s}(r) - \phi_0 s}{(1/2 - \phi_0) s } = \frac{2}{1 + \exp \left[r/l_D \right] }
\end{equation}
It is interesting to note that the specific dependence of this expression on $\phi_0$ is the same as for mosaic structures (see Appendix \ref{sec:mosaic}), although no such constraint was imposed in Ref.~\cite{Ma:2020} for the numerical reconstruction. The apparent average mean curvature, estimated from the derivative of this expression is two times smaller than Eq. (\ref{eq:Debye_H}). This difference is also clearly visible through the different slopes of the curves in Fig. \ref{fig:Debye_F0s} for small values of $r$.

\begin{figure}
\includegraphics[width=8cm]{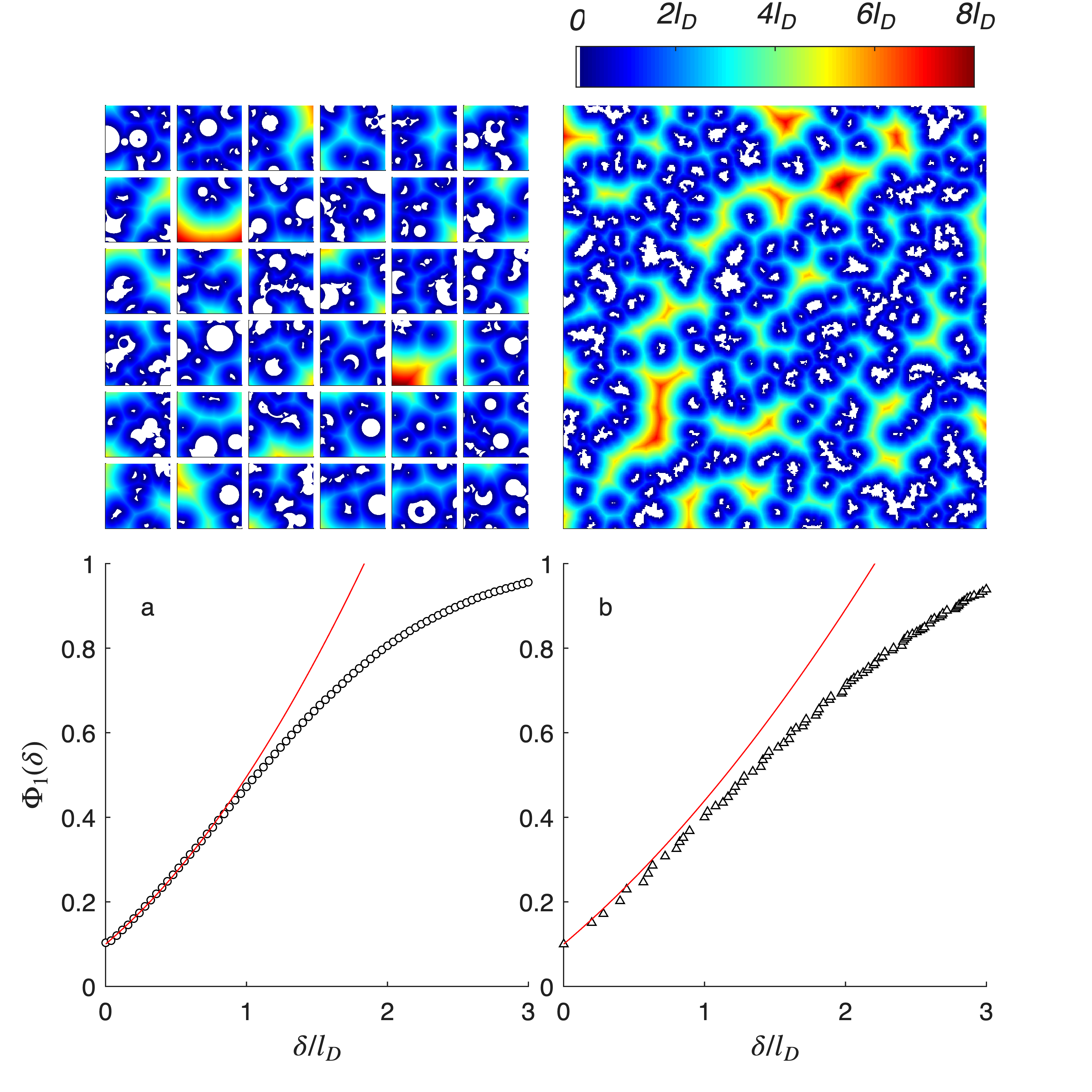}
\caption{Steiner-offset analysis of the surface curvature of dead-leaves structures with exponential covariance (left) and of numerically-reconstructed Debye random media (right, from Ref.~\cite{Ma:2020}). The colors in the upper figures represent the distance from any pore point to the closest surface; the bottom figures plot the volume fraction of points at distance smaller than $\delta$ from any surface. The dots are measured on the realizations, and the solid red line is Eq. (\ref{eq:Steiner}). }
\label{fig:Steiner}
\end{figure}

The dead-leaves structure with size distribution form Eq. (\ref{eq:FR_3}) and the numerically-reconstructed Debye random media are homometric structures \cite{Patterson:1939,Jiao:2010B,Gommes:2012,Skolnick:2021}: they have the same exact two-point correlation function and they can therefore not be discriminated by classical scattering experiments. It is a striking observation that two homometric structures can have average mean curvatures differing by a factor as large as two. To better understand the structural significance of that observation, Fig. \ref{fig:Steiner} explores a Steiner-offset approach for comparing the surface curvature of the dead-leaves and Debye structures. In the figure, two-dimensional cuts through 3D structures with $\phi_1=1$ are shown. The procedure consists in dilating the solid with a disk with radius $\delta$, and  measuring the solid fraction $\Phi_1(\delta)$ of the dilated set. This is equivalent to considering the set of all points at distance shorter than $\delta$ from the solid surface. For small distances $\delta$ the values obey Steiner's formula, namely \cite{Serra:1982} %\cite{Schamberger:2023}
\begin{equation} \label{eq:Steiner}
\Phi_1( \delta ) = \phi_1 + s_{2D} \left[ \delta  - \frac{1}{2} \bar H_{2D} \delta^2 + \ldots \right]
\end{equation}
where the two-dimensional surface area $s_{2D}$ and curvature $\bar H_{2D}$ are related to the three-dimensional values through the following stereological relations
\begin{equation}
s_{2D} = \frac{\pi}{4} s
\quad \textrm{and} \quad
\bar H_{2D} = \frac{1}{2} \bar H
\end{equation}
with $s= 4 \phi_0 \phi_1/l_D$. Fig. \ref{fig:Steiner}a shows that the numerical values of $\Phi_1(\delta)$ measured on the realizations of the dead-leaves model are nicely captured by Eq. (\ref{eq:Steiner}) with the curvature from Eq. (\ref{eq:Debye_H}). For the numerically-reconstructed Debye medium, the reconstruction domain has to extend over several $l_D$'s to capture the exponential two-point correlation function, and this necessarily limits the resolution. The data in Fig. \ref{fig:Steiner}b are nevertheless compatible with Eq. (\ref{eq:Steiner}), with a value of $\bar H$ two times smaller than that of the dead-leaves. 

The derivative $d \Phi_1(\delta)/d \delta$ of the curves shown in Fig. \ref{fig:Steiner} captures how the surface area of the dilated solid changes with the dilation distance $\delta$. From that perspective, the figure testifies to two opposing trends. On one hand the overall convexity of the solid is responsible for the bending upwards of the $\Phi_1(\delta)$ curves, which is very clear in the dead-leaves structure. On the other hand the dilation smoothes out the irregularities of the solid surface, which by itself bends the $\Phi_1(\delta)$ curves downwards. The latter effect is visible in the Debye structure of Fig. \ref{fig:Steiner}b. It therefore appears that the significantly lower value of $\bar H$ for the numerically-reconstructed Debye media, essentially points at a much rougher surface than in the dead-leaves structure. 
 
\begin{figure}
\includegraphics[width=7cm]{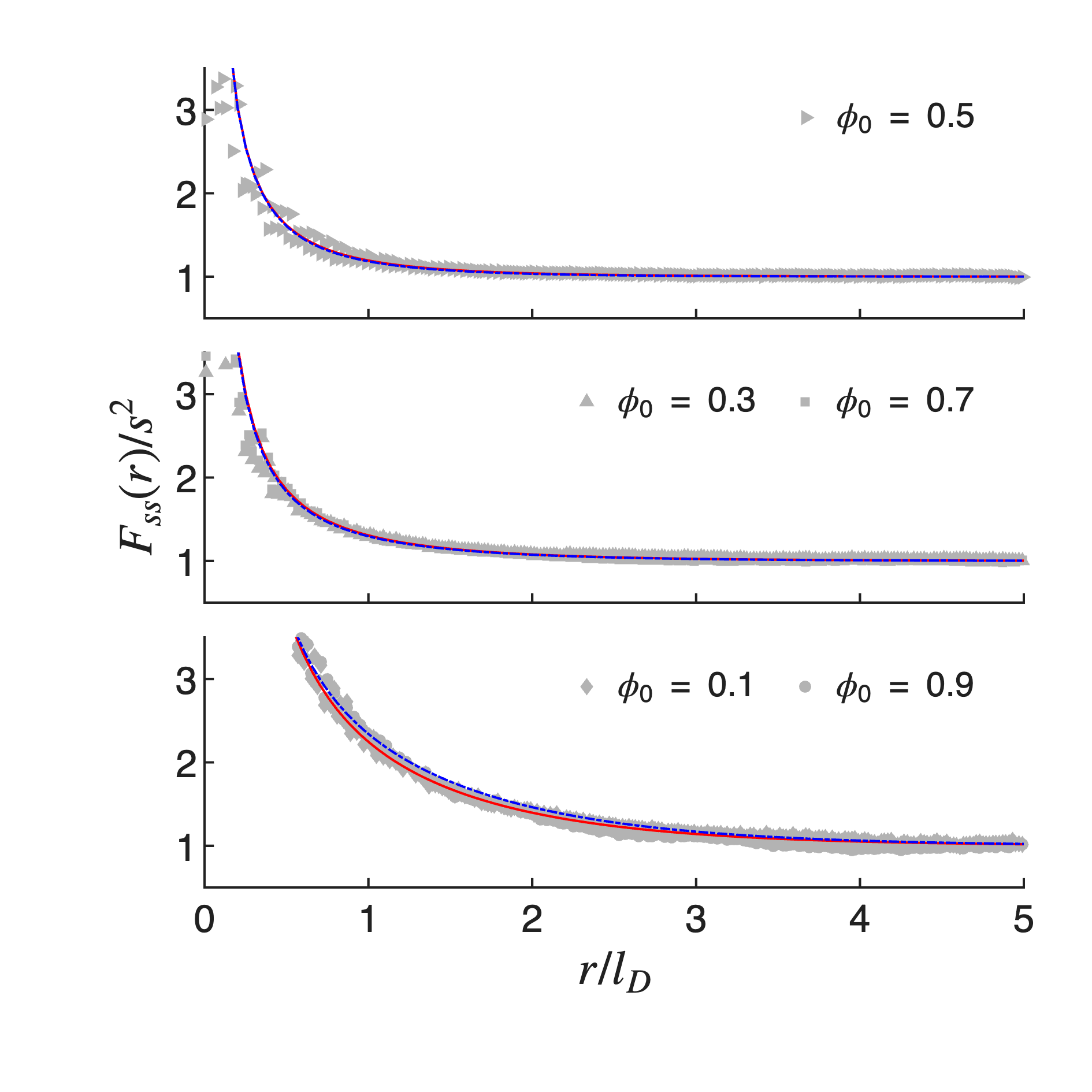}
\caption{Surface-surface correlation functions $F_{ss}(r)$ of Debye random media obtained as a dead-leaves model (solid red line) or via simulated annealing (dashed blue). The symbols are measured from realizations of the dead-leaves model.}
\label{fig:Debye_Fss}
\end{figure}

Interestingly, the surface-surface correlation function $F_{ss}(r)$ does not enable one to discriminate the dead-leaves structure with exponential covariance from numerically-reconstructed Debye media. The values are plotted in Fig. \ref{fig:Debye_Fss}, where the dots and red lines are for the dead-leaves model. The blue lines are obtained from the following semi-empirical expression
\begin{equation} \label{eq:Debye_Fss}
F_{ss}(r) = \frac{s}{2r} \exp \left[- r/l_D \right]  + s^2 + \frac{1}{l_D^2} \frac{|1/2 -\phi_0 |}{1 + \exp \left[r/l_D \right] }
\end{equation}
obtained in Ref.~\cite[Eq. (45)]{Ma:2020} from numerically reconstructed Debye media. Unlike the pore-surface correlation function, the values of $F_{ss}(r)$ obtained from the dead-leaves model or via numerical reconstruction are very similar. The data testify to small differences for small or large porosities ($\phi_0=0.1$ or $\phi_0 = 0.9$), but they may result from numerical uncertainties.

\section{Conclusions}

The main results of the paper are the general expressions for the pore-surface and surface-surface correlation functions of the dead-leaves model, $F_{0s}(r)$ and $F_{ss}(r)$, given in Eqs. (\ref{eq:deadleaves_F0s}) and (\ref{eq:deadleaves_Fss}). To derive these expressions, a recursive formalism was developed, wherein the properties of the dead-leaves model are obtained as stable points of iteration relations. This novel approach was validated by showing how it reproduces the classical expression of the two-point correlation function of the dead-leaves model. The new expressions of the surface correlation functions are further confirmed by numerical simulations.

The expressions of $F_{0s}(r)$ and $F_{ss}(r)$ are valid for arbitrary grains, in any dimension. All the geometrical characteristics of the individual grains that are relevant to the surface correlation functions of the dead-leaves structure, are shown to be captured by the generalized volume- and surface-covariograms $K_v(r)$, $K_{vs}(r)$ and $K_{ss}(r)$. These functions, sketched in Fig. \ref{fig:covariograms}, characterize the self- and cross-correlations between the volume and the surface of the individual grains. Interestingly, the known expressions for the surface correlation functions of the Boolean model of spheres (and disks) can also be expressed in terms of the generalized covariograms. We submit that the so-obtained expressions (in Eqs. \ref{eq:Boolean_F0s} and \ref{eq:Boolean_Fss}) are valid for arbitrary grains, and this is confirmed by numerical simulations for Boolean models of hollow spheres.

To illustrate the surface correlation functions of the dead-leaves model, we considered spherical grains having a specific size distribution, designed to yield a structure with exponential two-point correlation function. Because structures having that property have been explored by other authors using a general numerical reconstruction procedure, this provides a unique opportunity to investigate the degeneracy of two-point correlation functions. In particular, we show that the dead-leaves structure and numerically-reconstructed Debye media have distinctly different pore-surface correlation functions. The average mean curvature of the surface is two times smaller for the numerical reconstructions, which results from their extremely rough surface. The surface-surface correlation function does not seem to capture any significant difference between the two structures. 

\begin{acknowledgments}
The author is grateful to the Fonds de la Recherche Scientifique (F.R.S.-FNRS, Belgium) for a Research Associate Position.
\end{acknowledgments}

\appendix

\section{Pore-surface correlation function of mosaic structures}
\label{sec:mosaic}

We consider here a general mosaic structure, {\it i.e.} defined in two steps: (i) by creating first a tessellation, whereby space is decomposed into disconnected cells, followed (ii) by randomly and independently assigning each cell to the pores or the solid with probability $\phi_0$ and $\phi_1 = 1 -\phi_0$. 

For the purpose of calculating pore-surface correlation functions, we consider a two-phase structure obtained by dilating the boundaries of the tessellation by a sphere with radius $\epsilon/2$ (phase $B_\epsilon$), and eroding the cells by the same quantity (phase $C_\epsilon$). The volume fractions of the phases are
\begin{equation}  
\phi_{B_\epsilon} = s_t \epsilon + \mathcal{O}(\epsilon^2)
\end{equation}
and $\phi_{C_\epsilon} = 1 - \phi_{B_\epsilon}$ where $s_t$ is the specific surface area of the tessellation. This should not be confused with the surface area of the resulting mosaic, calculated as
\begin{equation} \label{eq:s_mosaic}
s = 2 \phi_0 \phi_1 s_t
\end{equation}
where the factor is the probability for two adjacent cells not to be both solid-like or pore-like.

The probability for two points at distance $r$ from one another to be in a cell and in the boundary, respectively, can be calculated as
\begin{equation}
C_{C_\epsilon B_\epsilon}(r) = \textrm{Prob} \left\{ \mathbf{x} \in B_\epsilon \right\} \times  \textrm{Prob}\left\{  \mathbf{x} + \mathbf{r} \in C_\epsilon |  \mathbf{x} \in B_\epsilon\right\}
\end{equation}
where the first probability is $\phi_{B_\epsilon}$. The second (conditional) probability is equal to 1 in the limit of small $\epsilon$, because the two-dimensional boundaries are a set with vanishing measure in three-dimensional space. As a consequence
\begin{equation} \label{eq:CCB}
C_{C_\epsilon B_\epsilon}(r) = s_t \epsilon + \mathcal{O}(\epsilon^2)
\end{equation}
where any $r$-dependence is of order $\epsilon^2$. 

We also define $\tilde C_{C_\epsilon B_\epsilon}(r)$ to be the probability for point $\mathbf{x}$ to be in $C_\epsilon$ and for $\mathbf{x} + \mathbf{r}$ to be in a boundary of the {\it same cell} as $\mathbf{x}$. Unlike $C_{C_\epsilon B_\epsilon}(r)$, the function $\tilde C_{C_\epsilon B_\epsilon}(r)$ is dependent on $r$, in a way that characterizes the size and shape of the individual cells. The dependence is of the type 
\begin{equation} \label{eq:CCB_tilde}
\tilde C_{C_\epsilon B_\epsilon}(r) = s_t \epsilon \times F_{CB}(r) + \mathcal{O}(\epsilon^2)
\end{equation}
where $F_{CB}(r)$ is a dimensionless function satisfying $F_{CB}(r \to 0) = 1$ because $\tilde C_{C_\epsilon B_\epsilon}(r) \simeq C_{C_\epsilon B_\epsilon}(r)$ when $r$ is smaller than a typical cell dimension, and $F_{CB}(r \to \infty) = 0$.

With these definitions, the two-point correlation function between the pores of the mosaic and its surface layer $S_\epsilon$, defined in Eq. (\ref{eq:C0S_def}), can be written as follows in terms of the underlying tessellation
\begin{equation} \label{eq:C0s_mosaic}
C_{0S_\epsilon}(r) = \phi_0 \phi_1 \tilde C_{C_\epsilon B_\epsilon}(r) + \phi_0 \times 2 \phi_0 \phi_1 \left[ C_{C_\epsilon B_\epsilon}(r)  - \tilde C_{C_\epsilon B_\epsilon}(r) \right]
\end{equation}
The first term accounts for the situation where $\mathbf{x}$ is in a cell and $\mathbf{x}+\mathbf{r}$ is on the surface of the same cell. This event contributes to $C_{0S_\epsilon}(r)$ with probability $\phi_0 \phi_1$, because the cell has to be pore-like and the adjacent cell solid-like. The second term accounts for the situation where $\mathbf{x}+\mathbf{r}$ is not on the surface of the same cell as $\mathbf{x}$. In that case, $\mathbf{x}$ is in a pore with probability $\phi_0$, and $\mathbf{x} + \mathbf{r}$ is on a surface of the mosaic with probability $2\phi_0 \phi_1$. The latter factor has the same meaning as in Eq. (\ref{eq:s_mosaic})

Using Eq. (\ref{eq:F0s_limit}) to calculate the pore-surface correlation function as a limit for $\epsilon \to 0$, the expression in Eq. (\ref{eq:C0s_mosaic}) can be written as follows
\begin{equation} \label{eq:mosaic_F0s}
\frac{F_{0s}(r) - \phi_0 s}{(1/2 - \phi_0) s} = F_{CB}(r)
\end{equation}
where we have used Eq. (\ref{eq:s_mosaic}), Eq. (\ref{eq:CCB}) and Eq. (\ref{eq:CCB_tilde}). Equation (\ref{eq:mosaic_F0s}) holds for any mosaic, and Eq. (\ref{eq:F0s_mosaic_DL}) is a particular case, which provides an explicit expression for $F_{CB}(r)$ in the case of the dead-leaves tessellation.

\section{Arguments for the validity of Eqs. (\ref{eq:Boolean_F0s}) and (\ref{eq:Boolean_Fss}) for Boolean models with arbitrary grains}
\label{sec:Boolean}

The expression for the two-point correlation function of the Boolean model in Eq. (\ref{eq:Boolean_C00}) is valid for any grain, in arbitrary dimension \cite{Matheron:1967,Serra:1982,Lantuejoul:2002}. All the characteristics of the grain relevant to $C_{00}(r)$ are therefore captured by the covariogram $K_v(r)$, defined in Eq. (\ref{eq:K_def}). In the case of the dead-leaves model, all the characteristics of the grains relevant to the surface correlation functions $F_{0s}(r)$ and $F_{ss}(r)$ are the covariograms $K_v(r)$, $K_{vs}(r)$ and $K_{ss}(r)$ as defined in Fig. \ref{fig:covariograms}. If we assume that the same holds for the Boolean model, then the expressions in Eqs. (\ref{eq:Boolean_F0s}) and (\ref{eq:Boolean_Fss}) should hold for any grain.

To support this idea, we first point out that equations (\ref{eq:Boolean_F0s}) and (\ref{eq:Boolean_Fss}) happen to be true also for two-dimensional disks. The covariograms of the disk with radius $R$,  are 
\begin{equation} \label{eq:Kv_D}
K_v^{(D)}(r) = 2 R^2 \left[ \cos^{-1}\left( \frac{r}{2R} \right) - \frac{r}{2R} \sqrt{1 - \left(\frac{r}{2R} \right)^2} \right] \Theta(2R-r)
\end{equation}
\begin{equation} \label{eq:Kvs_D}
K_{vs}^{(D)}(r) = 2 R \cos^{-1} \left(\frac{r}{2R} \right) \Theta(2R-r)
\end{equation}
and
\begin{equation} \label{eq:Kss_D}
K_{ss}^{(D)}(r) = \frac{2R}{r} \left[ 1-\left( \frac{r}{2R} \right)^2 \right]^{-1/2} \Theta(2R-r)
\end{equation}
With these expressions, the values of $F_{0s}(r)$ and $F_{ss}(r)$ in Ref. \cite{Ma:2020,Samarin:2023} are seen to be particular cases of Eqs. 
(\ref{eq:Boolean_F0s}) and (\ref{eq:Boolean_Fss}).

\begin{figure}
\includegraphics[width=12cm]{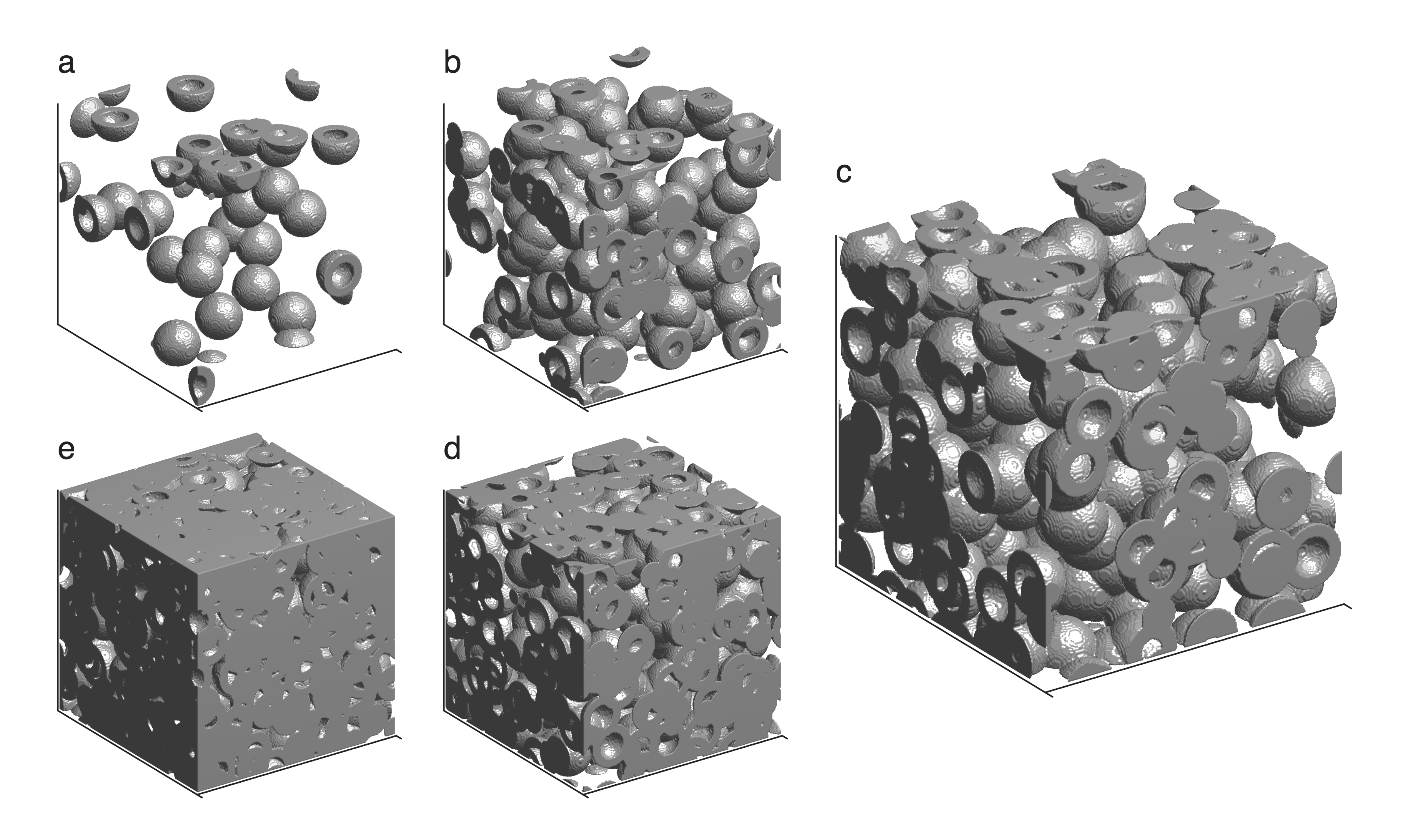}
\caption{Realizations of a Boolean model of interpenetrating hollow spheres, with pore volume fractions $\phi_0 = 0.1$ (a), 0.3 (b), 0.5 (c), 0.7 (d) and 0.9 (e).}
\label{fig:hollow_spheres}
\end{figure}

\begin{figure}
\includegraphics[width=7cm]{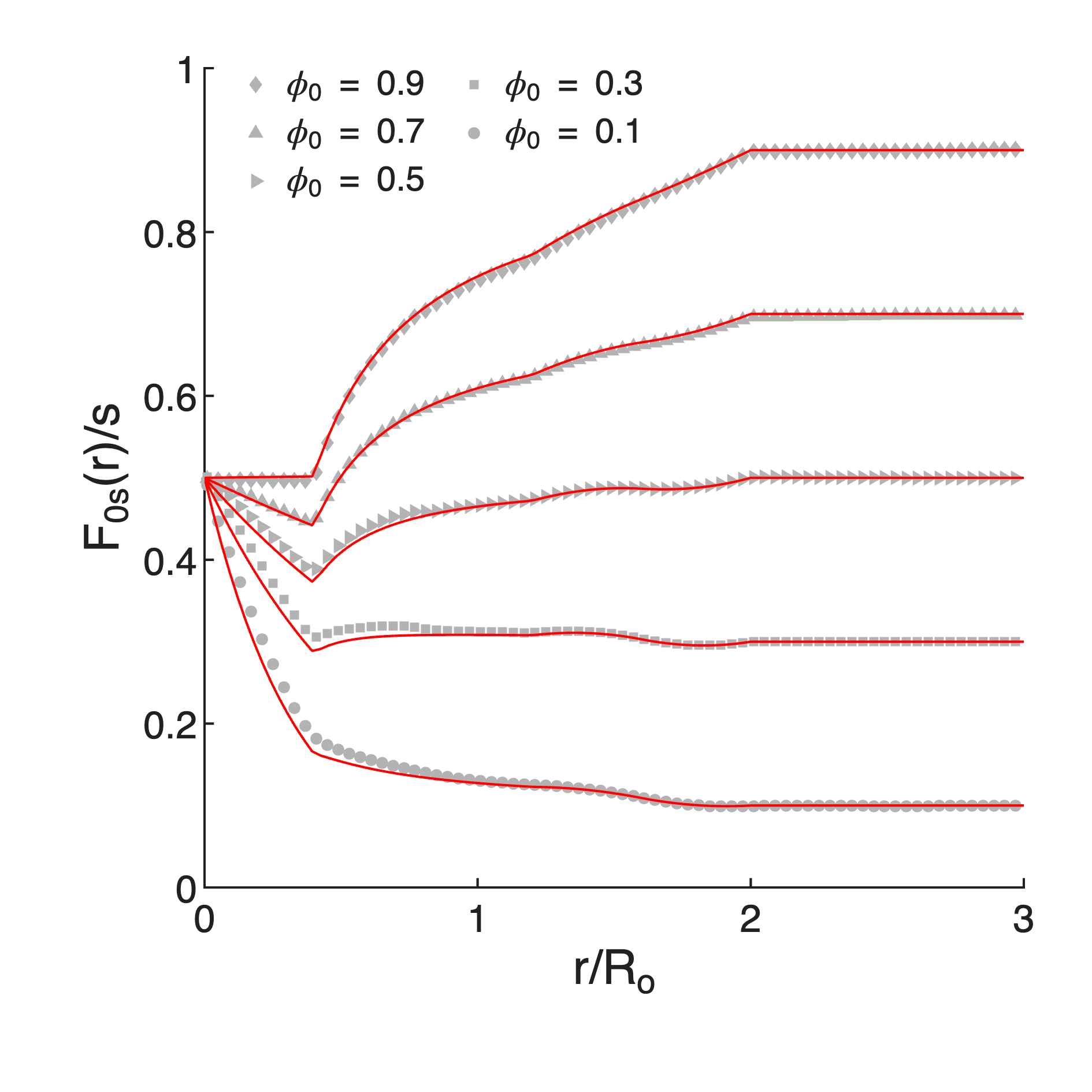}
\caption{Pore-surface correlation functions of the Boolean model of interpenetrating hollow spheres, with various pore volume fractions $\phi_0$. The symbols were measured of realizations similar to Fig. \ref{fig:hollow_spheres}, and the solid lines were calculated through Eq. (\ref{eq:Boolean_F0s}).}
\label{fig:F0s_hollow_spheres}
\end{figure}

\begin{figure}
\includegraphics[width=7cm]{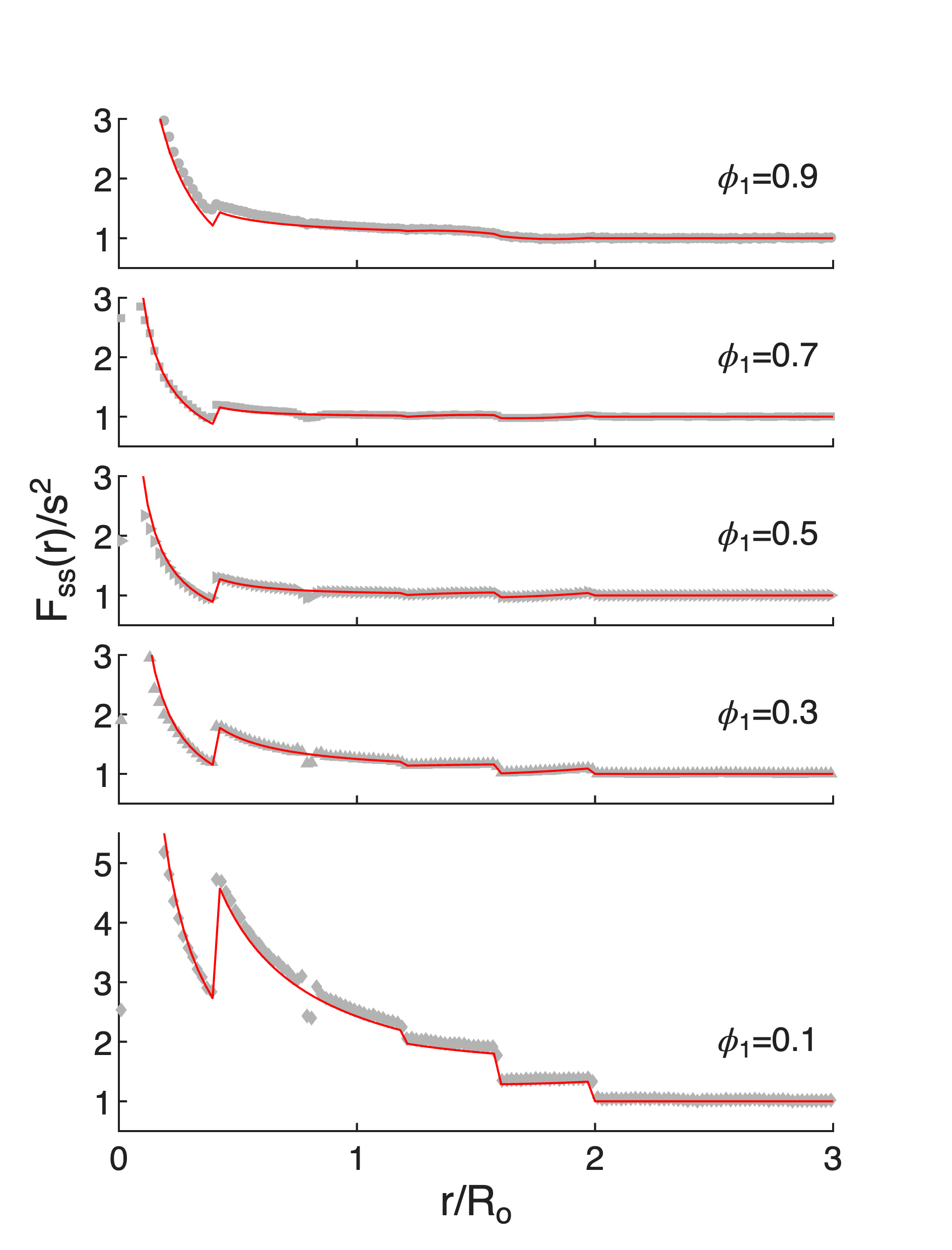}
\caption{Surface-surface correlation functions of the Boolean model of interpenetrating hollow spheres, with various pore volume fractions $\phi_0$. The symbols were measured of realizations similar to Fig. \ref{fig:hollow_spheres}, and the solid lines were calculated through Eq. (\ref{eq:Boolean_Fss}).}
\label{fig:Fss_hollow_spheres}
\end{figure}

For arbitrary grains, not necessarily spherical or circular, we pointed out in the main text that small-distance behavior of $F_{0s}(r)$ reproduced the average mean curvature of the individual grains in the limit of small solid fractions (see Eq. \ref{eq:Boolean_H}). This is indeed the expected behavior of $F_{0s}(r)$ of the Boolean model when the grains are too distant to overlap. To numerically explore the validity of Eqs. (\ref{eq:Boolean_F0s}) and (\ref{eq:Boolean_Fss}) for finite $r$ and arbitrary solid fraction, we consider here the case of a Boolean model of hollow spheres (see Fig. \ref{fig:hollow_spheres}). The numerically evaluated values of the pore-surface and surface-surface correlation functions $F_{0s}(r)$ and $F_{ss}(r)$ are plotted in Figs. \ref{fig:F0s_hollow_spheres} and \ref{fig:Fss_hollow_spheres}.

To compare the numerically evaluated values of the pore-surface correlation function $F_{0s}(r)$ with Eq. (\ref{eq:Boolean_F0s}), one needs only calculate the covariograms. For a hollow sphere with inner and outer radii $R_i$ and $R_o$, the covariogram $K_v(r)$ is calculated as 
\begin{equation} \label{eq:Kv_H}
K^{(H)}_v(r) = K_v^{(i)}(r) + K_v^{(o)} - 2 K_v^{(io)}(r)
\end{equation}
where $K_v^{(i)}(r)$ and $K_v^{(o)}$ are the covariograms of the inner and outer spheres, calculated from Eq. (\ref{eq:Kv_S}), and the last term is the intersection volume of two spheres with radii $R_i$ and $R_o$ at distance $r$ from one another. The latter is calculated as
\begin{eqnarray}
K_v^{(io)}(r) = \left\{
\begin{array}{lcl}
\frac{4 \pi}{3} R_i^3 & \textrm{for} & r < R_o-R_i  \cr
 \frac{4}{3} \left[ h_o^2 (3R_o-h_o) + h_i^2(3R_i-h_i) \right]  & \textrm{for} & R_o-R_i \leq r < R_o + R_i \cr
 0 & \textrm{for} & R_o + R_i \leq r 
\end{array}
\right.
\end{eqnarray}
where 
\begin{equation}
h_o = \frac{1}{2r} \left[ R_i^2 - (R_o-r)^2 \right]
\quad \textrm{and} \quad
h_i = \frac{1}{2r} \left[ R_o^2 - (R_i-r)^2 \right]
\end{equation}
are the heights of the two spherical caps, with radii $R_o$ and $R_i$, that make up the intersection of the two spheres.

The volume-surface covariogram $K_{vs}(r)$ of the hollow sphere is 
\begin{equation} \label{eq:Kvs_H}
K^{(H)}_{vs}(r) = K_{vs}^{(o)}(r) - K_{vs}^{(i)}(r) + K_{vs}^{(oi)}(r) - K_{vs}^{(io)}(r)
\end{equation}
where $K_{vs}^{(o)}(r)$ and $K_{vs}^{(i)}$ are the volume-surface covariograms of the inner and outer spheres, calculated from Eq. (\ref{eq:Kvs_S}). The third term, $K_{vs}^{(oi)}(r)$ results from intersecting the volume of the outer sphere and the surface of the inner sphere. It takes the value
\begin{eqnarray}
K_{vs}^{(oi)}(r) = \left\{
\begin{array}{lcl}
4 \pi R_i^2 & \textrm{for} & r < R_o-R_i  \cr
2 \pi R_i h_i & \textrm{for} & R_o-R_i \leq r < R_o + R_i \cr
 0 & \textrm{for} & R_o + R_i \leq r 
\end{array}
\right.
\end{eqnarray}
The fourth term, $K_{vs}^{(io)}(r)$ results from intersecting the volume of the inner sphere and the surface of the outer sphere. It takes the value
\begin{eqnarray}
K_{vs}^{(io)}(r) = \left\{
\begin{array}{lcl}
0 & \textrm{for} & r < R_o-R_i  \cr
2 \pi R_o h_o & \textrm{for} & R_o-R_i \leq r < R_o + R_i \cr
 0 & \textrm{for} & R_o + R_i \leq r 
\end{array}
\right.
\end{eqnarray}
The values of $F_{0s}(r)$ calculated from these expressions through Eq. (\ref{eq:Boolean_F0s}) are plotted in Fig. \ref{fig:F0s_hollow_spheres}. They nicely match the numerical values evaluated from the realizations, which supports the general validity of Eq. (\ref{eq:Boolean_F0s}) for arbitrary grains. 

To further compare the numerically evaluated values of the surface-surface correlation function $F_{ss}(r)$ with Eq. (\ref{eq:Boolean_Fss}), one needs to calculate the covariogram $K_{ss}(r)$ of the hollow sphere. The value is
\begin{equation} \label{eq:Kss_H}
K^{(H)}_{ss}(r) = K_{ss}^{(o)}(r) + K_{ss}^{(i)}(r) + 2 K_{ss}^{(oi)}(r)
\end{equation}
where the first two contributions are the surface-surface covariograms of the inner and outer spheres (calculated through Eq. \ref{eq:Kss_S}), and the last term results from the cross-correlation between the outer and inner surfaces. It is calculated as
\begin{equation}
K^{(oi)}_{ss}(r) = \frac{2 \pi R_i R_o}{r}  \quad \textrm{for} \quad R_o-R_i \leq r < R_o + R_i
\end{equation}
and $K^{(oi)}_{ss}(r) = 0$ outside of that range of $r$. The values of $F_{ss}(r)$ calculated from this expression through Eq. (\ref{eq:Boolean_Fss}) are plotted in Fig. \ref{fig:Fss_hollow_spheres}. They nicely match the numerical values evaluated from the realizations, which supports the general validity of Eq. (\ref{eq:Boolean_Fss}) for arbitrary grains.

\bibliography{deadleaves.bib}

\end{document}